\documentclass[12pt]{article}

\usepackage{amsmath,amssymb,bbm,latexsym,epsfig,graphicx,multirow}

\newcommand{\be}{\begin{equation}}
\newcommand{\ee}{\end{equation}}
\newcommand{\ba}{\begin{eqnarray}}
\newcommand{\ea}{\end{eqnarray}}
\newcommand{\bs}{\begin{subequations}}
\newcommand{\es}{\end{subequations}}
\newcommand{\no}{\nonumber\\}

\newcommand{\zz}{\mathbbm{Z}}
\newcommand{\mnu}{\mathcal{M}_\nu}
\newcommand{\diag}{\mbox{diag}}

\newcommand{\de}{\mathbf{10}}
\newcommand{\vi}{\mathbf{120}}
\newcommand{\se}{\mathbf{126}}
\newcommand{\seb}{\overline{\mathbf{126}}}
\newcommand{\du}{\mathbf{210}}
\newcommand{\qu}{\mathbf{45}}

\allowdisplaybreaks

\begin{document}

\title{
\normalsize \hfill UWThPh-2015-23
\\[.5mm]
\normalsize \hfill CFTP/15-010
\\[4mm]
\LARGE Flavour symmetries in a renormalizable $SO(10)$ model}

\author{
P.\ M.\ Ferreira,$^{(1,2)}$\thanks{E-mail: \tt pmmferreira@fc.ul.pt}
\addtocounter{footnote}{2}
W.~Grimus,$^{(3)}$\thanks{E-mail: \tt walter.grimus@univie.ac.at}
\
D.~Jur\v{c}iukonis,$^{(4)}$\thanks{E-mail: \tt darius.jurciukonis@tfai.vu.lt}
\ and
L.~Lavoura$^{(5)}$\thanks{E-mail: \tt balio@cftp.tecnico.ulisboa.pt}
\\*[3mm]
$^{(1)} \! $
\small ISEL---Instituto Superior de Engenharia de Lisboa, \\
\small Instituto Polit\'ecnico de Lisboa, Portugal
\\[2mm]
$^{(2)} \! $
\small Centro de F\'\i sica Te\'orica e Computacional, \\
\small Faculdade de Ci\^encias, Universidade de Lisboa, Portugal
\\[2mm]
$^{(3)} \! $
\small University of Vienna, Faculty of Physics, \\
\small Boltzmanngasse 5, A--1090 Wien, Austria
\\[2mm]
$^{(4)} \! $
\small University of Vilnius,
\small Institute of Theoretical Physics and Astronomy, \\
\small A.\ Go\v{s}tauto st.\ 12, Vilnius 01108, Lithuania 
\\[2mm]
$^{(5)} \! $
\small CFTP, Instituto Superior T\'ecnico, Universidade de Lisboa, \\
\small 1049-001 Lisboa, Portugal
\\*[2mm]
}

\date{22 March 2016}

\maketitle

\begin{abstract}
In the context of a renormalizable supersymmetric
$SO(10)$ Grand Unified Theory,
we consider the fermion mass matrices generated by the Yukawa couplings
to a $\mathbf{10} \oplus \mathbf{120} \oplus \overline{\mathbf{126}}$ 
representation of scalars.
We perform a complete investigation of the possibilities
of imposing flavour symmetries in this scenario;
the purpose is to reduce the number of Yukawa coupling constants
in order to identify
potentially
predictive models.
We have found that there are only 14 inequivalent cases
of Yukawa coupling matrices,
out of which 13 cases are generated by $\zz_n$ symmetries,
with suitable $n$,
and one case is generated by a $\zz_2 \times \zz_2$ symmetry.
A numerical analysis of the 14 cases reveals that
only two of them---dubbed A and B in the present paper---allow good fits
to the experimentally known fermion masses and mixings.
\end{abstract}

\newpage

\section{Introduction}

$SO(10)$ is a
popular gauge group
for the construction of Grand Unified Theories (GUTs).
The reason is that its 16-plet
accommodates
at once
all the chiral fields of one fermion family.
Now~\cite{slansky,ross},
\bs
\label{tensor}
\ba
\left( \mathbf{16} \otimes \mathbf{16} \right)_\mathrm{S}
&=& \mathbf{10} \oplus \mathbf{126},
\\
\left( \mathbf{16} \otimes \mathbf{16} \right)_\mathrm{AS}
&=& \mathbf{120},
\ea
\es
where the subscripts ``S'' and ``AS'' stand for,
respectively,
the symmetric and the antisymmetric parts of the tensor product.
Therefore,
in a renormalizable theory the scalars occurring in the Yukawa couplings
belong solely to the irreducible representations (irreps) $\mathbf{10}$,
$\overline{\mathbf{126}}$,
and $\mathbf{120}$.\footnote{The representations $\mathbf{10}$
and $\mathbf{120}$ are self-conjugate.}
Previously,
in the so-called
``minimal supersymmetric $SO(10)$
GUT'' (for an incomplete list of references see ref.~\cite{msgut,
Aulakh:2003kg,AG041}) the
$\mathbf{120}$ was absent.
However,
inconsistencies in the fit
of the experimental masses and mixings of the fermions---in particular,
a tension between the seesaw and GUT
scales~\cite{aulakh}---led to the inclusion of the 120-plet;
the resulting theory has been
called~\cite{nmsgut}
the ``new minimal supersymmetric $SO(10)$ GUT'' (NMSGUT)---see
ref.~\cite{oshimo}
and the references therein.\footnote{A completely different approach
is $SO(10)$ GUT models in extra dimensions---see for instance
ref.~\cite{feruglio}
and the references therein---or with a hidden sector~\cite{ludl}.} 

It has turned out that the NMSGUT,
which contains three 16-plets of fermionic fields
and one multiplet of scalars for each of the irreps
in the right-hand sides of equations~(\ref{tensor}),
is quite a successful theory and is capable of accommodating
all the available data on the fermion masses and mixings,
including the recent neutrino oscillation data~\cite{tortola,schwetz};
this has been demonstrated
by numerical fits~\cite{altarelli}.\footnote{Note that
skipping the $\overline{\mathbf{126}}$ of scalars does not allow for a good fit
of even the charged-fermion sector alone~\cite{LKG2006}.}
However,
adding a 120-plet to the 10-plet and the 126-plet of scalars
leads to a proliferation of parameters in the Yukawa couplings;
one might want to restrict the number of parameters in order to obtain
potentially predictive scenarios.
Attempts in this direction have been made:
in ref.~\cite{matsuda},
texture zeros were placed in the mass matrices;
in ref.~\cite{GK2006},
a $\zz_2$ flavour symmetry has been imposed together with a $CP$ symmetry;
in ref.~\cite{GK2007},
real Yukawa couplings were assumed
and $CP$ was broken solely by the imaginary vacuum expectation values (VEVs)
of the $\mathbf{120}$.

In the present paper we pursue the approach of ref.~\cite{GK2006}
by investigating all the possible flavour symmetries
acting on the Yukawa couplings in the NMSGUT.
We firstly perform a complete discussion by using only minimal assumptions;
we thereby identify all the possible cases and their
symmetry groups.
Thereafter,
all the cases
are subjected to a numerical analysis
in order to identify the viable ones.
Partially anticipating our results,
no non-Abelian flavour symmetry groups are permitted and
there are 14 inequivalent cases,
out of which 13
pertain to one-generator Abelian groups
and only one case has a two-generator symmetry group $\zz_2 \times \zz_2$.
However,
the numerical analysis rules out almost all the cases,
leaving only two viable ones which are compatible
with the data on the fermion masses and mixings.

In section~\ref{notation} we fix the notation,
display the basic formulas needed for our investigation,
and set forth our assumptions.
In section~\ref{cases} we list all the 14 cases.
The results of the numerical analysis are presented in section~\ref{numerics}.
The conclusions of our work are given in section~\ref{concl}.
The analysis of two specific problems that arise
in family symmetry-furnished GUTs is deferred to appendix~\ref{potential}. 
The discussion of the possibility of one further group generator
is left to appendix~\ref{investigation}.
Appendix~\ref{inequalities} focuses on the derivation of some inequalities
among the VEVs of the various $SO(10)$ scalar representations.

\section{Notation, framework, and assumptions}
\label{notation}

The relevant fermion mass matrices are given by 
(see for instance refs.~\cite{ross,senjanovic})
\bs
\label{mmm}
\ba
M_d &=& k_d\, H + \kappa_d\, G + v_d\, F, 
\label{md} \\
M_u &=& k_u\, H + \kappa_u\, G + v_u\, F, 
\label{mu} \\
M_\ell &=& k_d\, H + \kappa_\ell\, G - 3 v_d\, F, 
\label{ml} \\
M_D &=& k_u\, H + \kappa_D\, G - 3 v_u\, F,
\label{mD}
\ea
\es
where $M_d$,
$M_u$,
and $M_\ell$ are the mass matrices of the down-type quarks,
the up-type quarks,
and the charged leptons,
respectively,
while $M_D$ is the neutrino Dirac mass matrix.
The Yukawa-coupling matrices $H$,
$G$,
and $F$ are associated with the scalar irreps $\mathbf{10}$,
$\mathbf{120}$,
and $\overline{\mathbf{126}}$,
respectively.
Those matrices have the (anti)symmetry properties
\bs
\ba
H^T &=& H, \\
G^T &=& -G, \\
F^T &=& F.
\ea
\es
The coefficients $k_d$,
$v_d$,
$\kappa_d$,
and $\kappa_\ell$ are the VEVs of the Higgs doublet components
in the respective $SO(10)$ scalar irreps
which contribute to the Higgs doublet $H_d$
of the Minimal Supersymmetric Standard Model (MSSM).
The remaining coefficients---$k_u$,
$v_u$,
$\kappa_u$,
and $\kappa_D$---refer to $H_u$. 
The light-neutrino mass matrix is obtained as
\be
\label{mnu}
\mnu = M_L - M_D M_R^{-1} M_D^T
\ee
with
\bs
\ba
M_L &=& w_L\, F,
\\
M_R &=& w_R\, F,
\ea
\es
where $w_L$ and $w_R$ are the VEVs of scalar triplets
of the Pati--Salam~\cite{pati} group $SU(4)_c \times SU(2)_L \times SU(2)_R$,
which are part of the scalar 126-plet of $SO(10)$.
The first term in the right-hand side of equation~(\ref{mnu})
corresponds to the contribution
of the type~II seesaw mechanism~\cite{seesawII}
and the second term to the contribution
of the type~I seesaw mechanism~\cite{seesaw}.
Thus,
\be
\label{vioty}
\frac{w_R}{v_d}\, \mnu = \frac{w_L w_R}{v_d^2}\, M_d^F
- M_D \left( M_d^F \right)^{-1} \! M_D^T,
\ee
where $M_d^F \equiv v_d F$ is the component of the down-type-quark mass matrix
arising from the Yukawa coupling to the $\overline{\mathbf{126}}$ of scalars.
One sees that
\begin{itemize}
\item a complex factor $w_L w_R \left/ v_d^2 \right.$
parameterizes the strength of the type~II seesaw contribution
relative to the strength of the type~I seesaw contribution;
and
\item the overall magnitude of the neutrino masses
relative to the charged-fermion masses
is parameterized by a dimensionless factor $\left| w_R / v_d \right|$.
\end{itemize}
The mass Lagrangian of the ``light'' fermions reads 
\be
\mathcal{L}_\mathrm{mass} = 
- \bar d_L M_d d_R - \bar u_L M_u u_R - \bar \ell_L M_\ell \ell_R -
\frac{1}{2}\, \bar\nu_L \mnu \left( \nu_L \right)^c + \mbox{H.c.},
\ee
with $\left( \nu_L \right)^c = C \bar \nu_L^T$
being the charge-conjugate of $\nu_L$.
One diagonalizes the ``Hermitian mass matrices'' as
\bs
\ba
U_d^\dagger \left( M_d M_d^\dagger \right) U_d &=&
\mathrm{diag} \left( m_d^2,\ m_s^2,\ m_b^2 \right),
\\
U_u^\dagger \left( M_u M_u^\dagger \right) U_u &=&
\mathrm{diag} \left( m_u^2,\ m_c^2,\ m_t^2 \right),
\\
U_\ell^\dagger \left( M_\ell M_\ell^\dagger \right) U_\ell &=&
\mathrm{diag} \left( m_e^2,\ m_\mu^2,\ m_\tau^2 \right),
\\
U_\nu^\dagger \left( \mnu \mnu^\dagger \right) U_\nu &=&
\mathrm{diag} \left( m_1^2,\ m_2^2,\ m_3^2 \right),
\ea
\es
where the matrices $U_{d,u,\ell,\nu}$ are unitary
and $\left| m_3^2 - m_1^2 \right| \gg m_2^2 - m_1^2 > 0$.
The fermion mixing matrices are then
\bs
\ba
V \equiv U_\mathrm{CKM} &=& U_u^\dagger U_d,
\\
U_\mathrm{PMNS} &=& U_\ell^\dagger U_\nu.
\ea
\es
The neutrino mass spectrum is dubbed ``normal'' if $m_3^2 > m_1^2$
and ``inverted'' otherwise.

We make the following assumptions:
\begin{itemize}
\item All three matrices $H$, $F$, and $G$ are nonzero.
\item $\det F \neq 0$.
\item No generation decouples.
\end{itemize}
The second assumption is necessary for the type~I seesaw mechanism.
The third assumption is an experimental fact.

If the Lagrangian is invariant under a flavour symmetry $\mathcal{S}_0$,
then,
due to the $SO(10)$ structure of the Yukawa couplings
we obtain the following relations:
\be
\label{S0}
\mathcal{S}_0: \quad
\left\{ \begin{array}{rcl}
W^T H W e^{i\alpha} &=& H, \\*[1mm]
W^T G W e^{i\beta} &=& G, \\*[1mm]
W^T F W e^{i\gamma} &=& F,
\end{array} \right.
\ee
where $W$ is the $3 \times 3$ unitary matrix which acts on 
the three matter 16-plets under $\mathcal{S}_0$.
Without loss of generality we take $W$ to be diagonal.
The scalar multiplets $\mathbf{10}$,
$\mathbf{120}$,
and $\overline{\mathbf{126}}$ transform under $\mathcal{S}_0$
with the phase factors $e^{i\alpha}$,
$e^{i\beta}$,
and $e^{i\gamma}$,
respectively.
(One of the phase factors may be absorbed into $W$.)

\section{The 14 cases}
\label{cases}

\subsection{A single flavour symmetry}
\label{single}

A single symmetry transformation $\mathcal{S}_0$
leads to 13 inequivalent cases.
We refrain from going through the tedious arguments leading to these cases;
we merely list them instead.
In the following,
generic non-zero entries in the Yukawa coupling matrices
are denoted ``$\times$''. 
For each case,
we also give the Abelian group
through which the Yukawa-coupling matrices can be enforced.
\paragraph{Case A}
\bs\label{caseA}
\ba
& &
\label{S0-A}
\zz_2: \quad W = \diag \left( +1,\ +1,\ -1 \right), 
\quad e^{i\alpha} = +1,
\quad e^{i\beta} = -1,
\quad e^{i\gamma} = +1,
\\
& &
H \sim \left( \begin{array}{ccc}
\times & \times & 0 \\
\times & \times & 0 \\
0 & 0 & \times
\end{array} \right),
\quad
G \sim \left( \begin{array}{ccc}
0 & 0 & \times \\
0 & 0 & \times \\
\times & \times & 0
\end{array} \right),
\quad
F \sim \left( \begin{array}{ccc}
\times & \times & 0 \\
\times & \times & 0 \\
0 & 0 & \times
\end{array} \right).
\ea
\es
\paragraph{Case B}
\bs\label{caseB}
\ba
& & \zz_2: \quad 
W = \diag \left( +1,\ +1,\ -1 \right), 
\quad e^{i\alpha} = -1,
\quad e^{i\beta} = -1,
\quad e^{i\gamma} = +1,
\\
& &
H \sim \left( \begin{array}{ccc}
0 & 0 & \times \\
0 & 0 & \times \\
\times & \times & 0
\end{array} \right),
\quad
G \sim \left( \begin{array}{ccc}
0 & 0 & \times \\
0 & 0 & \times \\
\times & \times & 0
\end{array} \right),
\quad
F \sim \left( \begin{array}{ccc}
\times & \times & 0 \\
\times & \times & 0 \\
0 & 0 & \times
\end{array} \right).
\ea
\es
\paragraph{Case C}
\bs
\ba
\label{S0-C}
& & \zz_2: \quad W = \diag \left( +1,\ -1,\ +1 \right), 
\quad e^{i\alpha} = -1,
\quad e^{i\beta} = +1,
\quad e^{i\gamma} = +1,
\\
\label{HFG-C}
& &
H \sim \left( \begin{array}{ccc}
0 & \times & 0 \\
\times & 0 & \times \\
0 & \times & 0
\end{array} \right),
\quad
G \sim \left( \begin{array}{ccc}
0 & 0 & \times \\
0 & 0 & 0 \\
\times & 0 & 0
\end{array} \right),
\quad
F \sim \left( \begin{array}{ccc}
\times & 0 & \times \\
0 & \times & 0 \\
\times & 0 & \times
\end{array} \right).
\ea
\es
\paragraph{Case A$_1$}
\bs
\ba
& & \zz_4: \quad 
W = \diag \left( +1,\ -1,\ \pm i \right), 
\quad e^{i\alpha} = +1,
\quad e^{i\beta} = \mp i,
\quad e^{i\gamma} = -1,
\\ & &
H \sim \left( \begin{array}{ccc}
\times & 0 & 0 \\
0 & \times & 0 \\
0 & 0 & 0
\end{array} \right),
\quad
G \sim \left( \begin{array}{ccc}
0 & 0 & \times \\
0 & 0 & 0 \\
\times & 0 & 0
\end{array} \right),
\quad
F \sim \left( \begin{array}{ccc}
0 & \times & 0 \\
\times & 0 & 0 \\
0 & 0 & \times
\end{array} \right).
\ea
\es
\paragraph{Case A$^\prime_1$}
\bs
\ba
& &
U(1):
\quad W = \diag \left( 1,\ e^{2i\sigma},\ e^{i\sigma} \right),
\quad e^{i\alpha} = 1,
\quad e^{i\beta} = e^{-i\sigma},
\quad e^{i\gamma} = e^{-2i\sigma},
\\ & &
H \sim \left( \begin{array}{ccc}
\times & 0 & 0 \\
0 & 0 & 0 \\
0 & 0 & 0
\end{array} \right),
\quad
G \sim \left( \begin{array}{ccc}
0 & 0 & \times \\
0 & 0 & 0 \\
\times & 0 & 0
\end{array} \right),
\quad
F \sim \left( \begin{array}{ccc}
0 & \times & 0 \\
\times & 0 & 0 \\
0 & 0 & \times
\end{array} \right).
\ea
\es
\paragraph{Case A$^{\prime\prime}_1$}
\bs
\ba & &
U(1):
\quad W = \diag \left( e^{2i\sigma},\ 1,\ e^{i\sigma} \right),
\quad e^{i\alpha} = 1,
\quad e^{i\beta} = e^{-3i\sigma},
\quad e^{i\gamma} = e^{-2i\sigma},
\\ & &
H \sim \left( \begin{array}{ccc}
0 & 0 & 0 \\
0 & \times & 0 \\
0 & 0 & 0
\end{array} \right),
\quad
G \sim \left( \begin{array}{ccc}
0 & 0 & \times \\
0 & 0 & 0 \\
\times & 0 & 0
\end{array} \right),
\quad
F \sim \left( \begin{array}{ccc}
0 & \times & 0 \\
\times & 0 & 0 \\
0 & 0 & \times
\end{array} \right).
\ea
\es
\paragraph{Case A$_2$}
\bs
\ba & &
U(1):
\quad W = \diag \left( e^{i\sigma},\ e^{-i\sigma},\ 1 \right),
\quad e^{i\alpha} = 1,
\quad e^{i\beta} = e^{-i\sigma},
\quad e^{i\gamma} = 1,
\\ & &
H \sim \left( \begin{array}{ccc}
0 & \times  & 0 \\
\times  & 0 & 0 \\
0 & 0 & \times 
\end{array} \right),
\quad
G \sim \left( \begin{array}{ccc}
0 & 0 & \times \\
0 & 0 & 0 \\
\times & 0 & 0
\end{array} \right),
\quad
F \sim \left( \begin{array}{ccc}
0 & \times & 0 \\
\times & 0 & 0 \\
0 & 0 & \times
\end{array} \right).
\ea
\es
\paragraph{Case D$_1$}
\bs
\ba & &
\zz_3: \quad
W = \diag \left( \omega^2,\ \omega,\ 1 \right),
\quad e^{i\alpha} = 1,
\quad e^{i\beta} = \omega,
\quad e^{i\gamma} = \omega,
\label{iu1}
\\ & &
H \sim \left( \begin{array}{ccc}
0 & \times & 0 \\
\times & 0 & 0 \\
0 & 0 & \times
\end{array} \right),
\quad
G \sim \left( \begin{array}{ccc}
0 & 0 & \times \\
0 & 0 & 0 \\
\times & 0 & 0
\end{array} \right),
\quad
F \sim \left( \begin{array}{ccc}
0 & 0 & \times \\
0 & \times & 0 \\
\times & 0 & 0
\end{array} \right).
\ea
\es
\paragraph{Case D$_2$}
\bs
\ba & &
\zz_3:
\quad W = \diag \left( \omega,\ \omega^2,\ 1 \right),
\quad e^{i\alpha} = 1,
\quad e^{i\beta} = \omega^2,
\quad e^{i\gamma} = \omega,
\label{iu2}
\\ & &
H \sim \left( \begin{array}{ccc}
0 & \times & 0 \\
\times & 0 & 0 \\
0 & 0 & \times
\end{array} \right),
\quad
G \sim \left( \begin{array}{ccc}
0 & 0 & \times \\
0 & 0 & 0 \\
\times & 0 & 0
\end{array} \right),
\quad
F \sim \left( \begin{array}{ccc}
\times & 0 & 0 \\
0 & 0 & \times \\
0 &  \times & 0
\end{array} \right).
\ea
\es
\paragraph{Case D$_3$}
\bs
\ba & &
\zz_3:
\quad W = \diag \left( \omega,\ 1,\ \omega^2 \right),
\quad e^{i\alpha} = 1,
\quad e^{i\beta} = 1,
\quad e^{i\gamma} = \omega,
\label{iu3}
\\ & &
H \sim \left( \begin{array}{ccc}
0 & 0 & \times \\
0 & \times & 0 \\
\times & 0 & 0
\end{array} \right),
\quad
G \sim \left( \begin{array}{ccc}
0 & 0 & \times \\
0 & 0 & 0 \\
\times & 0 & 0
\end{array} \right),
\quad
F \sim \left( \begin{array}{ccc}
\times & 0 & 0 \\
0 & 0 & \times \\
0 &  \times & 0
\end{array} \right).
\ea
\es
\paragraph{Case D$^\prime_1$}
\bs
\ba & &
U(1):
\quad W = \diag \left( e^{-i\sigma},\ e^{i\sigma},\ e^{3i\sigma} \right),
\quad e^{i\alpha} = 1,
\quad e^{i\beta} = e^{-2i\sigma},
\quad e^{i\gamma} = e^{-2i\sigma},
\\ & &
H \sim \left( \begin{array}{ccc}
0 & \times & 0 \\
\times & 0 & 0 \\
0 & 0 & 0
\end{array} \right),
\quad
G \sim \left( \begin{array}{ccc}
0 & 0 & \times \\
0 & 0 & 0 \\
\times & 0 & 0
\end{array} \right),
\quad
F \sim \left( \begin{array}{ccc}
0 & 0 & \times \\
0 & \times & 0 \\
\times & 0 & 0
\end{array} \right).
\ea
\es
\paragraph{Case D$^\prime_2$}
\bs
\ba & &
U(1):
\quad W = \diag \left( e^{i\sigma},\ e^{-i\sigma},\ e^{3i\sigma} \right),
\quad e^{i\alpha} = 1,
\quad e^{i\beta} = e^{-4i\sigma},
\quad e^{i\gamma} = e^{-2i\sigma},
\\ & &
H \sim \left( \begin{array}{ccc}
0 & \times & 0 \\
\times & 0 & 0 \\
0 & 0 & 0
\end{array} \right),
\quad
G \sim \left( \begin{array}{ccc}
0 & 0 & \times \\
0 & 0 & 0 \\
\times & 0 & 0
\end{array} \right),
\quad
F \sim \left( \begin{array}{ccc}
\times & 0 & 0 \\
0 & 0 & \times \\
0 &  \times & 0
\end{array} \right).
\ea
\es
\paragraph{Case D$^\prime_3$}
\bs
\ba & &
U(1):
\quad W = \diag \left( e^{i\sigma},\ e^{3i\sigma},\ e^{-i\sigma} \right),
\quad e^{i\alpha} = 1,
\quad e^{i\beta} = 1,
\quad e^{i\gamma} = e^{-2i\sigma},
\\ & &
H \sim \left( \begin{array}{ccc}
0 & 0 & \times \\
0 & 0 & 0 \\
\times & 0 & 0
\end{array} \right),
\quad
G \sim \left( \begin{array}{ccc}
0 & 0 & \times \\
0 & 0 & 0 \\
\times & 0 & 0
\end{array} \right),
\quad
F \sim \left( \begin{array}{ccc}
\times & 0 & 0 \\
0 & 0 & \times \\
0 &  \times & 0
\end{array} \right).
\ea
\es
In equations~(\ref{iu1}),
(\ref{iu2}),
and~(\ref{iu3}) $\omega \equiv \exp{\left( \pm i 2 \pi / 3 \right)}$.

We note that only case~A had been discussed earlier,
in ref.~\cite{GK2006}.
Cases A$_1$ and A$_2$ have Yukawa-coupling matrices
which are restrictions
(\textit{i.e.}~they contain extra zero matrix elements)
of those of case A;
cases A$^\prime_1$ and A$^{\prime\prime}_1$ have Yukawa-coupling matrices
which are more restrictive than those of case A$_1$.

We demonstrate in appendix~\ref{potential} that the scalar potential of the
NMSGUT can consistently be modified in order to incorporate the $\zz_2$
symmetries present in cases~A, B and~C.

\subsection{A second flavour symmetry}

The list of 13 cases in the previous subsection 
does not necessarily comprise all the Yukawa-coupling matrices
obtainable through flavour symmetries,
because in each of those 13 cases
either one or more further symmetry transformations
might be operative and lead to more restrictive Yukawa-coupling matrices
and thus to new cases.
Let us denote a generic further symmetry transformation,
different from $\mathcal{S}_0$ of equation~(\ref{S0}),
by $\mathcal{S}_1$:
\be
\label{S1}
\mathcal{S}_1: \quad
\left\{ \begin{array}{rcl}
X^T H X e^{i\alpha_1} &=& H, \\*[1mm]
X^T G X e^{i\beta_1} &=& G, \\*[1mm]
X^T F X e^{i\gamma_1} &=& F.
\end{array} \right.
\ee
In principle,
the symmetry $\mathcal{S}_1$ might
either commute or not commute with $\mathcal{S}_0$.
However,
as shown in appendix~\ref{investigation},
by using our assumptions of section~\ref{notation}
one may demonstrate that $X$ always commutes with $W$,
\textit{i.e.}\ that $\mathcal{S}_1$ commutes with $\mathcal{S}_0$.
Even more surprisingly,
only one new case ensues,
which we denote by the letter~E
and is a subcase of both case~A and case~C:\footnote{It is also
a subcase of case~B,
as can be seen when one interchanges the first and third generations
in the matrices of equations~(\ref{caseB}).}
\paragraph{Case E}
\bs
\label{A3}
\ba
& &
\zz_2^{(1)}: \quad W = \diag \left( +1,\ +1,\ -1 \right), 
\quad e^{i\alpha} = +1,
\quad e^{i\beta} = -1,
\quad e^{i\gamma} = +1,
\\ & &
\zz_2^{(2)}: \quad X = \diag \left( +1,\ -1,\ +1 \right), 
\quad e^{i\alpha_1} = -1,
\quad e^{i\beta_1} = +1,
\quad e^{i\gamma_1} = +1,
\\*[1mm]
& &
H \sim \left( \begin{array}{ccc}
0 & \times & 0 \\
\times & 0 & 0 \\
0 & 0 & 0
\end{array} \right),
\quad
G \sim \left( \begin{array}{ccc}
0 & 0 & \times \\
0 & 0 & 0 \\
\times & 0 & 0
\end{array} \right),
\quad
F \sim \left( \begin{array}{ccc}
\times & 0 & 0 \\
0 & \times & 0 \\
0 & 0 & \times
\end{array} \right).
\ea
\es
Note that $\zz_2^{(1)}$ is the symmetry~(\ref{S0-A}) of case~A
while $\zz_2^{(2)}$ is the symmetry~(\ref{S0-C}) of case~C.

There are no possible cases for a flavour group with three or more generators.

\subsection{Summary}

From the assumptions stated in section~\ref{notation}
we have obtained the following results:
\begin{itemize}
\item There are 14 inequivalent cases.
\item All the cases except E 
can be obtained from a single flavour symmetry transformation.
\item The flavour groups with one generator
are the cyclic groups $\zz_2$ (in the cases~A, B, and~C),
$\zz_3$ (in the cases~D$_k$ with $k=1,2,3$),
and $\zz_4$ (in case~A$_1$).
The remaining cases have a $U(1)$ symmetry.\footnote{This $U(1)$
must be broken explicitly by the scalar potential,
which we did not consider here,
lest a Goldstone boson arises.
Therefore,
a full model will have a suitable cyclic symmetry group instead of $U(1)$.}
\item In case E there are two symmetry transformations
which commute with each other;
the flavour group is $\zz_2 \times \zz_2$.
\item Our scenario does not admit non-Abelian flavour groups.
\end{itemize}

\section{Fitting the cases to the data}
\label{numerics}

In this section we report on our numerical study of
cases~A, B,
C,
A$_1$,
and~D$_k$ ($k = 1, 2, 3$).
We have not studied the cases~A$^\prime_1$ and~A$^{\prime\prime}_1$
because they are restrictions of case~A$_1$
and we have found that that case is unable to fit the data well
(details will be given later).
Analogously,
the cases~D$^\prime_k$ are restrictions of the cases~D$_k$;
since we have found that the cases~D$_k$ do not work well,
we did not need to bother with the cases~D$^\prime_k$.
Finally,
case~E is a restriction of case~C
(and also of cases~A and~B);
since case~C is unable even to correctly fit the charged-fermion masses,
case~E can be discarded outright.

We did not attempt to fit case~A$_2$
because we knew beforehand that such an attempt would be unsuccessful.
Indeed,
case A$_2$ yields $M_d$ and $M_u$ of the Fritzsch form~\cite{fritzsch},
which has long been known to be unable to simultaneously fit the quark masses
and the CKM matrix.

\subsection{Parameter counting}

In order to get a feeling for the ability for fitting the data
that each case ought to have,
it is instructive to count the number of parameters
in each of the cases---see table~\ref{table-parameters}.
\begin{table}[ht!]
\centering
\begin{tabular}{|c||c|c|c|c|}
\hline
cases & A & B & C & A$_1$, D$_k$
\\ \hline
\# parameters in the & 13 moduli & 11 moduli & 10 moduli & 9 moduli
\\
$M_x M_x^\dagger$ for $x = d, \ell, u$ &
and 10 phases & and 7 phases & and 6 phases & and 5 phases
\\ \hline
\# extra parameters & 3 moduli & 3 moduli & 3 moduli & 3 moduli
\\
in $\mnu \mnu^\dagger$ & and 2 phases & and 2 phases & and 2 phases &
and 2 phases
\\ \hline
\end{tabular}
\caption{The number of parameters
in the Hermitian mass matrices for each case.}
\label{table-parameters}
\end{table}
For instance,
in case~A$_1$
the charged-fermion mass matrices may be written,
after adequate rephasings,
\bs
\ba
M_d &=& \left( \begin{array}{ccc}
a & 0 & f e^{i \theta_2} \\ 0 & c e^{i \theta_1} & b \\ - f e^{i \theta_2} & b & d
\end{array} \right),
\\
M_\ell &=& \left( \begin{array}{ccc}
3 a & 0 & g e^{i \theta_5} \\ 0 & c e^{i \theta_1} & 3 b \\ - g e^{i \theta_5} & 3 b & d
\end{array} \right),
\\
M_u &=& \left( \begin{array}{ccc}
t a & 0 & l e^{i \theta_4} \\ 0 & r c e^{i \left( \theta_1 + \theta_3 \right)} & t b \\
- l e^{i \theta_4} & t b & r d e^{i \theta_3}
\end{array} \right),
\ea
\es
with five phases $\theta_{1,2,3,4,5}$ and nine real and non-negative parameters
(``moduli'')
$a$,
$b$,
$c$,
$d$,
$f$,
$g$,
$l$,
$t \equiv \left| v_u / v_d \right|$,
and $r \equiv \left| k_u / k_d \right|$.
Moreover,
the neutrino mass matrix is
\be
\mnu = \left| \frac{v_d}{w_R} \right| \left( \begin{array}{ccc}
C a & - \left( rch / b \right) e^{i \left( \theta_1 + \theta_3 \right)} & 0 \\
- \left( rch / b \right) e^{i \left( \theta_1 + \theta_3 \right)} &
6 rct e^{i \left( \theta_1 + \theta_3 \right)} &
C b - \left( r^2cd / b \right) e^{i \left( \theta_1 + 2 \theta_3 \right)} \\
0 & C b - \left( r^2cd / b \right) e^{i \left( \theta_1 + 2 \theta_3 \right)} &
6 r t d e^{i \theta_3} - h^2 / a
\end{array} \right),
\ee
\textit{viz.}~it contains two extra complex parameters $C$ and $h$,
plus the real parameter $\left| v_d / w_R \right|$,
making an extra three moduli and two phases.

One sees in table~\ref{table-parameters} that
$\mnu \mnu^\dagger$ always contains three moduli and two phases
beyond the parameters which appear
in the charged-fermion Hermitian mass matrices.
It is easy to understand the reasons for that:
one extra complex parameter
originates in $\kappa_D$ of equation~(\ref{mD});
another complex parameter originates
in $w_L w_R \left/ v_d^2 \right.$ in the right-hand side
of equation~(\ref{vioty});
and there is an extra modulus $\left| w_R / v_d \right|$
in the left-hand side of equation~(\ref{vioty}).\footnote{Note that
the overall phase of $\mnu$ is unphysical.}

Case~A is the one that has most parameters,
hence most degrees of freedom,
in the mass matrices.
In ref.~\cite{GK2006} that case has been numerically studied
under some restrictive assumptions;
we have repeated that study under the same restrictive assumptions,
but using the updated values for the charged-fermion masses
given in ref.~\cite{zhou}.

The restriction of case~A analyzed in ref.~\cite{GK2006}
contains 13 moduli and 6 phases in the $M_x M_x^\dagger$ ($x = d, \ell, u$),
plus an extra two moduli and one phase in $\mnu \mnu^\dagger$.
The original ``minimal supersymmetric $SO(10)$ GUT''~\cite{msgut}
has 11 moduli and 8 phases in the $M_x M_x^\dagger$,
plus an extra two moduli and one phase in $\mnu \mnu^\dagger$.
We see that both those models are comparable to our case~B
in their numbers of parameters.

The $M_x M_x^\dagger$ are supposed to be able to fit 13 observables:
the nine charged-fermion masses and the four observables in the CKM matrix.
One must take into account that
phases usually do not help much in fitting observables;
the moduli are most relevant.
Additionally,
if one also takes into account $\mnu \mnu^\dagger$,
then we have to fit five parameters more---the three lepton mixing angles,
the ratio $r^2_\mathrm{solar} \equiv \left( m_2^2 - m_1^2 \right)
\left/ \, \left| m_3^2 - m_1^2 \right| \right.$,
and $\left| m_3^2 - m_1^2 \right|$ itself.
We have used the fixed value
$\left| m_3^2 - m_1^2 \right| = 2.5 \times 10^{-15}\, \mathrm{MeV}^2$,
which just allows us to determine the overall scale of $\mnu$,
\textit{viz.}~$\left| v_d / w_R \right|$.

\subsection{$\chi^2$ function}

In order to test the viability of each case,
and to find adequate numerical values for its parameters, 
we construct a $\chi^2$ function
\be
\chi^2 \left( x \right) = \sum_{i=1}^{n}
\left\{
H \left[ f_i \left( x \right) - \bar{O}_i \right]
\left( \frac{f_i \left( x \right) - \bar{O}_i}{\delta_+ O_i} \right)^2
+
H \left[ \bar{O}_i - f_i \left( x \right) \right]
\left( \frac{\bar{O}_i - f_i \left( x \right)}{\delta_- O_i} \right)^2
\right\},
\label{chi}
\ee
where $n$ is the total number of observables
(masses and mixing parameters)
to be fitted.
In equation~(\ref{chi}),
$H$ is the Heaviside step function,
$\bar{O}_i$ is the central value of each observable $O_i$,
$\delta_\pm O_i$ are the upper and lower errors of that observable,
and $f_i \left( x \right)$ is the value of that observable,
in any given case,
when the parameters of that case have the values $x = \{ x_\alpha \}$. 
The data are fitted by minimizing $\chi^2 \left( x \right)$
with respect to the $x_{\alpha}$.

We have used the mean values $\bar{O}$ and the errors $\delta_\pm O$
given in tables~\ref{table-masses}--\ref{table-nu}.
\begin{table}[ht!]
\renewcommand{\arraystretch}{1.25}
\centering
\begin{tabular}{|c||c|c|c|}
\hline \hline 
observable &
$m_d \left/ \, \mathrm{MeV} \right.$ &
$m_s \left/ \, \mathrm{MeV} \right.$ &
$m_b \left/ \, \mathrm{MeV} \right.$
\\
${\bar O}^{+ \delta_+ O}_{- \delta_- O}$ &
$0.70^{+0.31}_{-0.30}$ &
$13^{+4}_{-4}$ &
$790^{+40}_{-40}$
\\*[1mm] \hline \hline 
observable &
$m_e \left/ \, \mathrm{MeV} \right.$ &
$m_\mu \left/ \, \mathrm{MeV} \right.$ &
$m_\tau \left/ \, \mathrm{MeV} \right.$
\\
${\bar O}^{+ \delta_+ O}_{- \delta_- O}$ &
$0.283755495^{+2.4 \times 10^{-8}}_{-2.5 \times 10^{-8}}$ &
$59.9033617^{+5.4 \times 10^{-6}}_{-5.4 \times 10^{-6}}$ &
$1021.95^{+0.11}_{-0.12}$
\\*[1mm] \hline \hline 
observable &
$m_u \left/ \, \mathrm{MeV} \right.$ &
$m_c \left/ \, \mathrm{MeV} \right.$ &
$m_t \left/ \, \mathrm{MeV} \right.$
\\
${\bar O}^{+ \delta_+ O}_{- \delta_- O}$ &
$0.49^{+0.20}_{-0.17}$ &
$236^{+37}_{-36}$ &
$92200^{+9600}_{-7800}$
\\*[1mm] \hline \hline
\end{tabular}
\caption{The values of the charged-fermion masses used in our fits.}
\label{table-masses}
\end{table}
\begin{table}[h!]
\renewcommand{\arraystretch}{1.25}
\centering
\begin{tabular}{|c||c|c|c|c|}
\hline \hline 
observable &
$\left| V_{12} \right|$ &
$\left| V_{13} \right|$ &
$\left| V_{23} \right|$ &
$10^5\, J$
\\
${\bar O}^{+ \delta_+ O}_{- \delta_- O}$ &
$0.22536^{+0.00183}_{-0.00183}$ &
$0.00355^{+0.00045}_{-0.00045}$ &
$0.0414^{+0.0036}_{- 0.0036}$ & $3.06^{+0.63}_{-0.60}$
\\*[1mm] \hline \hline
\end{tabular}
\caption{The values of the CKM-matrix observables used in our fits.}
\label{table-CKM}
\end{table}
\begin{table}[ht!]
\renewcommand{\arraystretch}{1.25}
\centering
\begin{tabular}{|c||c|c|c|c|}
\hline \hline
observable &
$r^2_{\mathrm{solar}} \,(\mathrm{NH})$ &
$\sin^2\theta_{12} \,(\mathrm{NH})$ &
$\sin^2\theta_{13} \,(\mathrm{NH})$ &
$\sin^2\theta_{23} \,(\mathrm{NH})$
\\
${\bar O}^{+ \delta_+ O}_{- \delta_- O}$ &
$0.0306^{+0.0050}_{-0.0038}$ &
$0.323^{+0.052}_{-0.045}$ &
$0.0234^{+0.0060}_{-0.0057}$ &
$0.567^{+0.076}_{-0.175}$
\\*[1mm] \hline \hline 
observable &
$r^2_{\mathrm{solar}} \,(\mathrm{IH})$ &
$\sin^2\theta_{12} \,(\mathrm{IH})$ &
$\sin^2\theta_{13} \,(\mathrm{IH})$ &
$\sin^2\theta_{23} \,(\mathrm{IH})$
\\
${\bar O}^{+ \delta_+ O}_{- \delta_- O}$ &
$0.0319^{+0.0053}_{-0.0039}$ &
$0.323^{+0.052}_{-0.045}$ &
$0.0240^{+0.0057}_{-0.0057}$ &
$0.573^{+0.067}_{-0.172}$
\\*[1mm] \hline \hline
\end{tabular}
\caption{The values of the neutrino and lepton-mixing observables
used in our fits.
``NH''refers to a normal neutrino mass spectrum and ``IH'' to an inverted one.}
\label{table-nu}
\end{table}
We have taken
the charged-fermion masses in table~\ref{table-masses},
which are renormalized at $M_\mathrm{GUT} = 2 \times 10^{16}\, \mathrm{GeV}$,
from the last column of table V
of ref.~\cite{zhou}.\footnote{For other determinations
of the values of the running quark and lepton masses,
evolved from the electroweak scale to the GUT scale
through the renormalization group of the MSSM,
see ref.~\cite{das}.}
These are values computed using the renormalization-group equations
of the MSSM with $\tan{\beta} = 10$;
we leave it for some later,
more detailed study the task of fitting the data
for other values of $\tan{\beta}$.
The values of the CKM mixing angles in table~\ref{table-CKM}
are low-energy values and
were taken from equation~(12.27) of ref.~\cite{beringer};
we have multiplied the error bars given in that equation by a factor of three
in order to obtain adequately large intervals.
The values in table~\ref{table-nu} are the $3 \sigma$ intervals
given for each observable in ref.~\cite{tortola}.

In order to assess the fitting ability of each case,
we have firstly attempted to fit only the charged-fermion masses
(nine observables, given in table~\ref{table-masses}),
secondly the charged-fermion masses together with the CKM matrix
(four more observables,
given in table~\ref{table-CKM}),
and,
finally,
all that together with the neutrino masses and the PMNS matrix
(four observables more, given in table~\ref{table-nu}).
The total $\chi^2$ function is thus the sum of three terms:
\be
\chi^2_\mathrm{total} = \chi^2_\mathrm{masses} + \chi^2_\mathrm{CKM}
+ \chi^2_\nu.
\label{chi_total}
\ee
For the neutrino masses,
we have analysed both possibilities
of a normal or inverted neutrino mass spectrum;
indeed,
for each set of values for the parameters $x$,
we have computed the eigenvalues of $\mathcal{M}_\nu \mathcal{M}_\nu^\ast$
and thereby determined the type of neutrino mass spectrum;
we have then chosen accordingly the input values
in the computation of the function $\chi^2_\nu$.

In some cases we have not been able
to find a reasonably small value of $\chi^2_\mathrm{masses}$ alone;
in those cases,
further analysis by considering $\chi^2_\mathrm{CKM}$ and $\chi^2_\nu$ 
made no sense.
Similarly,
in some other cases a sufficiently low value
of $\chi^2_\mathrm{masses} + \chi^2_\mathrm{CKM}$
could not be achieved,
so we did not have to consider $\chi^2_\nu$.
Finally,
even when $\chi^2_\mathrm{total}$ could be correctly fitted,
we still had to check whether $\left| w_R / v_d \right|$ turned out
in the right range.
Indeed,
since $v_d$ must be of order the Fermi scale $100\, \mathrm{GeV}$
and $w_R$ must be of order the grand-unification scale $10^{16}\, \mathrm{GeV}$,
we must require $\left| w_R / v_d \right|$ to be $10^{14}$ or even larger.
We had to check some other inequalities,
the exposition of which we defer to section~\ref{B}.

\subsection{Numerical method}

The minimization of $\chi^2 \left( x \right)$
is a difficult task because the various parameters $x_\alpha$
may differ by several orders of magnitude
and because there always is a large number of local minima.
We have spent much time in the numerical analysis
trying to find absolute minima;
this has involved various fitting options
and restrictions of the parameters for each particular case.
Still,
we cannot be 100\% sure that
we have found the absolute minimum for all cases---the possibility remains
that a better solution exists somewhere in parameter space. 

For the numerical minimization of the $\chi^2$ functions
we have employed the Differential Evolution (DE) algorithm.
This is a stochastic algorithm
that exploits a population of potential solutions
in order to effectively probe the parameter space.
It was first introduced in ref.~\cite{storn}
and it has been modified several times since then. 

The effectiveness of the DE algorithm strongly depends on control parameters. 
We have performed preliminary tests
in order to hand-tune the appropriate ranges for the control parameters
in each case.
Also,
in the $\chi^2$ function of equation~(\ref{chi}),
we have modified the errors $\delta_\pm O_i$ randomly 
(within the range of magnitude of the true errors)
according to the behaviour of the fits;
we have thus been able to test,
for each case,
more local minima---defined as the points where
the minimization algorithm converges---and to find the minima
closer to the global minimum. 

All the numerical calculations were implemented
by using the programming language {\tt Fortran}.

\subsection{Case~B} \label{B}

\subsubsection{Theoretical treatment}

We choose a weak basis in which the 
Yukawa-coupling matrix $F$ is diagonal.
After an interchange of the first and third generations,
\be
\label{bkpt}
k_d H = \left( \begin{array}{ccc}
0 & d & h \\ d & 0 & 0 \\ h & 0 & 0
\end{array} \right),
\quad
\kappa_d G = \left( \begin{array}{ccc}
0 & f & g \\ -f & 0 & 0 \\ -g & 0 & 0
\end{array} \right),
\quad
v_d F = \left( \begin{array}{ccc}
a & 0 & 0 \\ 0 & b & 0 \\ 0 & 0 & c
\end{array} \right).
\ee
Without loss of generality, we assume the parameters $a$, $b$,
and $c$ to be non-negative real.
Then the mass matrices are given by 
\bs \label{w} \ba
M_d &=& \left( \begin{array}{ccc}
a & d+f & h+g \\
d-f & b & 0 \\
h-g & 0 & c
\end{array} \right),
\\
M_\ell &=& \left( \begin{array}{ccc}
-3a & d + \left( \kappa_\ell / \kappa_d \right) f &
h + \left( \kappa_\ell / \kappa_d \right) g \\
d - \left( \kappa_\ell / \kappa_d \right) f & -3b & 0 \\
h - \left( \kappa_\ell / \kappa_d \right) g & 0 & -3c
\end{array} \right),
\\
M_u &=& \left( \begin{array}{ccc}
\left( v_u / v_d \right) a &
\left( k_u / k_d \right) d + \left( \kappa_u / \kappa_d \right) f &
\left( k_u / k_d \right) h + \left( \kappa_u / \kappa_d \right) g \\
\left( k_u / k_d \right) d - \left( \kappa_u / \kappa_d \right) f &
\left( v_u / v_d \right) b &
0 \\
\left( k_u / k_d \right) h - \left( \kappa_u / \kappa_d \right) g &
0 &
\left( v_u / v_d \right) c
\end{array} \right),
\nonumber \\ & &
\\
M_D &=& \left( \begin{array}{ccc}
-3 \left( v_u / v_d \right) a &
\left( k_u / k_d \right) d + \left( \kappa_D / \kappa_d \right) f &
\left( k_u / k_d \right) h + \left( \kappa_D / \kappa_d \right) g \\
\left( k_u / k_d \right) d - \left( \kappa_D / \kappa_d \right) f &
-3 \left( v_u / v_d \right) b & 0 \\
\left( k_u / k_d \right) h - \left( \kappa_D / \kappa_d \right) g & 0 &
-3 \left( v_u / v_d \right) c
\end{array} \right).
\nonumber \\ & &
\label{mdw}
\ea
\es
We rewrite the mass matrices~(\ref{w}) as
\bs \label{ww} \ba
M_d &=& \left( \begin{array}{ccc}
a & k_1 & k_3 \\
k_2 & b & 0 \\
k_4 & 0 & c
\end{array} \right),
\label{Md} \\
M_\ell &=& \left( \begin{array}{ccc}
-3a & k_5 & k_7 \\
k_6 & -3b & 0 \\
k_8 & 0 & -3c
\end{array} \right),
\label{Mell} \\
M_u &=& \left( \begin{array}{ccc}
ta & k_9 & k_{11} \\
k_{10} & tb & 0 \\
k_{12} & 0 & tc
\end{array} \right),
\\
M_D &=& \left( \begin{array}{ccc}
-3ta & k_{13} & k_{15} \\
k_{14} & -3tb & 0 \\
k_{16} & 0 & -3tc
\end{array} \right),
\label{prp}
\ea
\es
where $t \equiv v_u / v_d$.
The $k_{1,2,\ldots,16}$ are not all independent.
We choose $k_{1,2,3,4,5,9,10,13}$ as parameters,
while
\bs
\label{jfipr}
\ba
k_6 &=& k_1 + k_2 - k_5, \\
k_7 &=& \frac{k_1 k_4 + k_3 k_5 - k_2 k_3 - k_4 k_5}{k_1 - k_2}, \\
k_8 &=& \frac{k_1 k_3 + k_4 k_5 - k_2 k_4 - k_3 k_5}{k_1 - k_2}, \\
k_{11} &=& \frac{\left( k_1 k_3 - k_2 k_4 \right) k_9
+ \left( k_1 k_4 - k_2 k_3 \right) k_{10}}
{k_1^2 - k_2^2}, \\
k_{12} &=& \frac{\left( k_1 k_3 - k_2 k_4 \right) k_{10}
+ \left( k_1 k_4 - k_2 k_3\right) k_9}
{k_1^2 - k_2^2}, \\
k_{14} &=& k_9 + k_{10} - k_{13}, \\
k_{15} &=& \frac{k_{13} \left( k_3 - k_4 \right)}{k_1 - k_2}
+ \frac{\left( k_9 + k_{10} \right) \left( k_1 k_4 - k_2 k_3 \right)}
{k_1^2 - k_2^2}, \\
k_{16} &=& \frac{k_{13} \left( k_4 - k_3 \right)}{k_1 - k_2}
+ \frac{\left( k_9 + k_{10} \right) \left( k_1 k_3 - k_2 k_4 \right)}
{k_1^2 - k_2^2}.
\ea
\es
From equations~(\ref{vioty}),
(\ref{bkpt}),
and~(\ref{prp}) it is easy to compute
\ba
\frac{w_R}{v_d}\, \mnu &=& \left( \frac{w_L w_R}{v_d^2} - 9 t^2 \right)
\left( \begin{array}{ccc} a & 0 & 0 \\ 0 & b & 0 \\ 0 & 0 & c
\end{array} \right)
\nonumber \\ & &
-
\left( \begin{array}{ccc}
\left. k_{13}^2 \right/ \! b + \left. k_{15}^2 \right/ \! c &
- 3 t \left( k_{13} + k_{14} \right) &
- 3 t \left( k_{15} + k_{16} \right) \\*[1mm]
- 3 t \left( k_{13} + k_{14} \right) &
\left. k_{14}^2 \right/ \! a &
\left. k_{14} k_{16} \right/ \! a \\*[1mm]
- 3 t \left( k_{15} + k_{16} \right) &
\left. k_{14} k_{16} \right/ \! a &
\left. k_{16}^2 \right/ \! a
\end{array} \right).
\ea

Next,
we multiply $M_u$ and $\mnu$ by
phase factors $\exp{\left( - i \arg{t} \right)}$, defining 
\bs
\ba
M_u^\prime &\equiv& \exp{\left( - i \arg{t} \right)}\, M_u,
\\
\mnu^\prime &\equiv& \exp{\left( - 2 i \arg{t} \right)}\, \mnu.
\ea
\es
This phase change leads to the redefinitions
\be
k^\prime_p \equiv k_p\,
\exp{\left( -i \arg{t} \right)}
\quad \mbox{for}\ p = 9, \ldots, 16.
\ee
Crucially,
equations~(\ref{jfipr}) remain valid
when using the $k^\prime_p$ instead of the $k_p$ for $p = 9, \ldots, 16$.
One obtains
\bs
\label{fbipt}
\ba
M_u^\prime &=& \left( \begin{array}{ccc}
\left| t \right| a & k_9^\prime & k_{11}^\prime \\
k_{10}^\prime & \left| t \right| b & 0 \\
k_{12}^\prime & 0 & \left| t \right| c
\end{array} \right),
\\
\frac{w_R}{v_d}\, \mnu^\prime &=&
\left( \hat C - 9 \left| t \right|^2 \right)
\left( \begin{array}{ccc}
a & 0 & 0 \\
0 & b & 0 \\ 
0 & 0 & c 
\end{array} \right)
\nonumber \\ & &
-
\left( \begin{array}{ccc}
\left. {k_{13}^\prime}^2 \right/ \! \left| b \right|
+ \left. {k_{15}^\prime}^2 \right/ \! \left| c \right| &
- 3 \left| t \right| \left( k_{13}^\prime + k_{14}^\prime \right) &
- 3 \left| t \right| \left( k_{15}^\prime + k_{16}^\prime \right) \\*[1mm]
- 3 \left| t \right| \left( k_{13}^\prime + k_{14}^\prime \right) &
\left. {k_{14}^\prime}^2 \right/ \! \left| a \right| &
\left. k_{14}^\prime k_{16}^\prime \right/ \! \left| a \right| \\*[1mm]
- 3 \left| t \right| \left( k_{15}^\prime + k_{16}^\prime \right) &
\left. k_{14}^\prime k_{16}^\prime \right/ \! \left| a \right| &
\left. {k_{16}^\prime}^2 \right/ \! \left| a \right|
\end{array} \right), \hspace*{5mm}
\label{gytuy}
\ea
\es
where $\hat C \equiv \left( w_L w_R / v_d^2 \right)
\exp{\left( - 2 i \arg{t} \right)}$.

In equations~(\ref{Md}), (\ref{Mell}), and~(\ref{fbipt}) 
one observes that the mass matrices of case~B
may be parameterized through five real quantities $a$,
$b$,
$c$,
$|t|$,
and $\left| w_R / v_d \right|$,
plus nine complex parameters 
$k_{1,2,3,4,5}$, 
$k^\prime_{9,10,13}$, 
and $\hat C$.
This justifies the third column of table~\ref{table-parameters}.

\subsubsection{Inequalities} \label{ineq}

We fix the scale $\left| w_R / v_d \right|$
in the left-hand side of equation~(\ref{gytuy})
by requiring the difference of the squared neutrino masses
$\left| m_3^2 - m_1^2 \right|$
to be equal to the atmospheric mass scale $2.5 \times 10^{-3}\, \mathrm{eV}^2$.
Afterwards,
we compute $\left| v_d \right|$
by identifying $\left| w_R \right|$ with the unification scale
$M_\mathrm{GUT} = 2 \times 10^{16}\, \mathrm{GeV}$.
Finally,
we calculate $\left| v_u \right| = \left| t v_d \right|$
from the value of the parameter $\left| t \right|$ of the fit.

In the supersymmetric GUT that we envisage there is only one
scalar doublet with hypercharge $+1/2$,
\textit{viz.}~$H_d$,
at the Fermi mass scale;
there is also only one scalar doublet with hypercharge $-1/2$,
\textit{viz.}~$H_u$,
at that scale.
Those two doublets have VEVs
\be
\left\langle H_d^0 \right\rangle_0
= \frac{174\, \mathrm{GeV}}{\sqrt{1 + \tan^2{\beta}}}
\quad \mbox{and} \quad
\left\langle H_u^0 \right\rangle_0
= \frac{\left( 174\, \mathrm{GeV} \right) \tan{\beta}}
{\sqrt{1 + \tan^2{\beta}}},
\ee
respectively,
where $\tan{\beta} = 10$ in our fit.
According to the inequalities~\eqref{vhpty},
\bs
\label{guiyo}
\ba
\left( \left\langle H_d^0 \right\rangle_0 \right)^2
&\ge&
\left| v_d \right|^2 + \left| k_d \right|^2 + \left| \kappa_d \right|^2
+ \frac{1}{3} \left| \kappa_\ell \right|^2,
\\
\left( \left\langle H_u^0 \right\rangle_0 \right)^2
&\ge&
\left| v_u \right|^2 + \left| k_u \right|^2 + \left| \kappa_u \right|^2
+ \frac{1}{3} \left| \kappa_D \right|^2.
\ea
\es
Therefore,
we have first of all to enforce the inequalities
\bs
\label{phyud}
\ba
\frac{\left( 2 \times 10^{16}\, \mathrm{GeV} \right)^2}
{\left| w_R / v_d \right|^2} &<&
\frac{\left( 174\, \mathrm{GeV} \right)^2}{1 + \tan^2{\beta}},
\label{pfort} \\
\left| t \right|^2 \frac{\left( 2 \times 10^{16}\, \mathrm{GeV} \right)^2}
{\left| w_R / v_d \right|^2} &<&
\frac{\left( 174\, \mathrm{GeV} \right)^2 \tan^2{\beta}}{1 + \tan^2{\beta}},
\ea
\es
on our fits.
The inequality~(\ref{pfort}) reads
$\left| w_R / v_d \right| > 1.155 \times 10^{15}$,
which is a quite useful lower bound.

Our fit fixes
\bs
\ba
\frac{\kappa_\ell}{\kappa_d} &=& \frac{k_5 - k_6}{k_1 - k_2},
\\
\frac{\kappa_u}{\kappa_d} &=& \frac{k_9 - k_{10}}{k_1 - k_2},
\\
\frac{\kappa_D}{\kappa_d} &=& \frac{k_{13} - k_{14}}{k_1 - k_2},
\\
\frac{k_u}{k_d} &=& \frac{k_9 + k_{10}}{k_1 + k_2}.
\ea
\es
Therefore,
from the inequalities~(\ref{guiyo}),
\bs
\label{glpsy}
\ba
\left| k_d \right|^2 + \left( 1 + \frac{1}{3}
\left| \frac{k_5 - k_6}{k_1 - k_2} \right|^2
\right) \left| \kappa_d \right|^2 &\le&
\left( \left\langle H_d^0 \right\rangle_0 \right)^2 - \left| v_d \right|^2,
\\
\left| \frac{k_9^\prime + k_{10}^\prime}{k_1 + k_2} \right|^2 \left| k_d \right|^2
+ \frac{\left| k_9^\prime - k_{10}^\prime \right|^2
+ \left( 1/3 \right)
\left| k_{13}^\prime - k_{14}^\prime \right|^2}{\left| k_1 - k_2 \right|^2}\,
\left| \kappa_d \right|^2 &\le&
\left( \left\langle H_u^0 \right\rangle_0 \right)^2 - \left| v_u \right|^2.
\ea
\es

We must now face the additional fact that the Yukawa couplings
cannot be too large,
lest the theory ceases to be perturbative
and/or Landau poles arise in the Yukawa couplings.
Let $y > 0$ denote the maximum value that we accept
for the absolute value of the Yukawa couplings;
we may take $y$ somewhere between 1 and 10.
From equations~(\ref{bkpt}),
\bs
\label{sioup}
\ba
\mathrm{max} \left( \left| d \right|, \left| h \right| \right)
&<& \left| k_d \right| y,
\\
\mathrm{max} \left( \left| f \right|, \left| g \right| \right)
&<& \left| \kappa_d \right| y,
\\
\mathrm{max} \left( a, b, c \right)
&<& \left| v_d \right| y. \label{hbuiyp}
\ea
\es
We have directly enforced the inequality~(\ref{hbuiyp}) on our fits.
The other two inequalities~\eqref{sioup}
may be put together with the inequalities~\eqref{glpsy} to derive
\bs
\label{iupwu}
\ba
\mathrm{max} \left( \left| k_1 + k_2 \right|^2,
\left| k_3 + k_4 \right|^2 \right)
& & \no
+ \left( 1
+ \frac{1}{3} \left| \frac{k_5 - k_6}{k_1 - k_2} \right|^2
\right)
\mathrm{max} \left( \left| k_1 - k_2 \right|^2,
\left| k_3 - k_4 \right|^2 \right)
&\le&
4 \left[
\left( \left\langle H_d^0 \right\rangle_0 \right)^2 - \left| v_d \right|^2
\right] y^2,
\\
\left| \frac{k_9^\prime + k_{10}^\prime}{k_1 + k_2} \right|^2
\mathrm{max} \left( \left| k_1 + k_2 \right|^2,
\left| k_3 + k_4 \right|^2 \right)
+ \left( \left| k_9^\prime - k_{10}^\prime \right|^2
\right. & & \no
\left.
+ \frac{1}{3} \left| k_{13}^\prime - k_{14}^\prime \right|^2
\right)
\frac{\mathrm{max} \left( \left| k_1 - k_2 \right|^2,
\left| k_3 - k_4 \right|^2 \right)}{\left| k_1 - k_2 \right|^2}
&\le&
4 \left[
\left( \left\langle H_u^0 \right\rangle_0 \right)^2 - \left| v_u \right|^2
\right] y^2. \hspace*{10mm}
\label{rutym}
\ea
\es

To summarize,
we have enforced on our fits the inequalities~\eqref{phyud},
\eqref{hbuiyp},
and~\eqref{iupwu}.

\subsubsection{Fit}

We have found that case~B is able to fit
\emph{perfectly}\/ all the observables.
This is true irrespective of whether
the neutrino mass spectrum is normal or inverted.
However,
when the neutrino mass spectrum is inverted some of the inequalities
in the previous subsection always turn out to be violated;
this happens because either $\left| w_R / v_d \right| < 10^{15}$ is too small
or $\left| t \right| > 300$ is so large that the inequality~\eqref{rutym}
ends up being violated.

For a normal neutrino mass spectrum,
on the other hand,
there are fits in which all the inequalities are observed.
In table~\ref{tableBnormal} we give the values of the mass-matrix parameters
that lead to the best fit which we have been able to achieve.
\begin{table}[ht!]
\renewcommand{\arraystretch}{1.25}
\centering
\begin{tabular}{|c||c|}
\hline
parameter & value \\ \hline \hline
$a \left/ \, \mathrm{MeV} \right.$ & 219.850545793720272 \\ \hline
$b \left/ \, \mathrm{MeV} \right.$ & 0.561919252512016 \\ \hline
$c \left/ \, \mathrm{MeV} \right.$ & 28.64031991278612 \\ \hline
$\left| t \right|$ & 1.558686846443802 \\ \hline
$k_1 \left/ \, \mathrm{MeV} \right.$ &
$106.613768172192835\, \exp{\left( i\, 2.726661945096518 \right)}$ \\ \hline
$k_2 \left/ \, \mathrm{MeV} \right.$ &
$3.214360308597388\, \exp{\left( i\, 5.73665831290545 \right)}$ \\ \hline
$k_3 \left/ \, \mathrm{MeV} \right.$ &
$750.563494049026872\, \exp{\left( i\, 4.735747016077402 \right)}$ \\ \hline
$k_4 \left/ \, \mathrm{MeV} \right.$ &
$20.603622366818627\, \exp{\left( i\, 1.445003798120007 \right)}$ \\ \hline
$k_5 \left/ \, \mathrm{MeV} \right.$ &
$10.49565466331894\, \exp{\left( i\, 4.813783368633092 \right)}$ \\ \hline
$k_9^\prime \left/ \, \mathrm{MeV} \right.$ &
$12964.825027004273579\, \exp{\left( i\, 3.963802982387908 \right)}$ \\ \hline
$k_{10}^\prime \left/ \, \mathrm{MeV} \right.$ &
$19.353350623796356\, \exp{\left( i\, 5.971066682817606 \right)}$ \\ \hline
$k_{13}^\prime \left/ \, \mathrm{MeV} \right.$ &
$22682.297777225823666\, \exp{\left( i\, 5.075844560968867 \right)}$ \\ \hline
$\hat C$ & $3291007.008905897848\, \exp{\left( i\, 2.868704037387841 \right)}$
\\ \hline
$\left| w_R / v_d \right|$ & $1.67257 \times 10^{15}$  \\ \hline
\end{tabular}
\caption{The values of the parameters for out best fit of case~B.}
\label{tableBnormal}
\end{table}
The value of $\chi^2_\mathrm{total}$ for this fit is smaller than $10^{-3}$,
\textit{i.e.},
for all practical purposes,
it is zero.
The smallest neutrino mass for this fit
is $m_1 \approx 0.006\, \mathrm{eV}$,
while $m_1 + m_2 + m_3 \approx 0.07\, \mathrm{eV}$.

It is interesting to observe
in table~\ref{tableBnormal} that the best fit is achieved
for a very large value of $\left| \hat C \right| \sim 10^6$,
meaning that the type-II seesaw mechanism dominates over the type-I.
For this fit, the matrices $U_d$ and $U_u$ are almost diagonal,
with $U_d$ mostly identical with $U_\mathrm{CKM}$.
On the other hand, $U_\nu$ is almost completely a rotation
between the first two generations, while $U_\ell$ is largely,
but not exclusively, a rotation between the second and third
generations; both rotations are almost maximal.

Since very perfect fits can be obtained in case~B,
we suspect that this case has too many degrees of freedom
and has little or no predictive power.
However,
since such a study is very time-consuming,
we leave it for later investigation.

\subsection{Non-viable cases}

We have found that all the cases except cases~A and~B
either fail to fit the observables adequately
or give a much too low value for $\left| w_R / v_d \right|$.
(For us,
an acceptable fit is one in which
all the observables simultaneously are within their ranges
in tables~\ref{table-masses}--\ref{table-nu}.)
Indeed,
case~C even fails to adequately fit the charged-fermion masses alone,
while cases~A$_1$,
D$_2$,
and~D$_3$ are unable to acceptably fit
the charged-fermion masses together with the CKM matrix.
The best results that we were able to find for all the cases
are given in table~\ref{table-fails}.
\begin{table}[ht!]
\renewcommand{\arraystretch}{1.25}
\centering
\begin{tabular}{|c||c|c|c|}
\hline
\multirow{2}{*}{case} &
\multirow{2}{*}{$\chi^2$ of best fit} &
pulls larger than one &
\multirow{2}{*}{remarks}
\\
& & in absolute value &
\\ \hline \hline
\multirow{8}{*}{A$_1$} & $\chi^2_{\mathrm{masses}} \sim 10^{-6}$ & &
\\*[1mm] \cline{2-4} 
 & \multirow{3}{*}
{$\chi^2_\mathrm{masses} + \chi^2_\mathrm{CKM} = 11.57$} &
$m_d: -1.83$ & \\
 & & $m_s: -1.49$ & \\
 & & $m_b: +2.17$ & \\*[1mm]
\cline{2-4}
 & \multirow{4}{*}{$\chi^2_\mathrm{total} = 19.26$} &
$m_d: -1.79$ & \multirow{4}{*}{normal hierarchy} \\
 & & $m_s: -1.44$ & \\
 & & $m_b: +2.27$ & \\ 
 & & $\sin^2{\theta_{23}}: -2.28$ & \\ \hline \hline
\multirow{3}{*}{D$_2$} &
\multirow{1}{*}{$\chi^2_\mathrm{masses} = 3.21$} &
$m_d: -1.74$ & \\ \cline{2-4}
 &
\multirow{2}{*}{$\chi^2_\mathrm{masses} + \chi^2_\mathrm{CKM} = 12.75$} &
$m_d: -2.09$ & \\ 
 & & $m_b: +2.54$ & \\*[1mm] \hline \hline
\multirow{3}{*}{D$_3$} & $\chi^2_\mathrm{masses} \sim 10^{-6}$ & &
\\*[1mm] \cline{2-4} 
 &
\multirow{2}{*}{$\chi^2_\mathrm{masses} + \chi^2_\mathrm{CKM} = 11.86$} &
$m_d: -1.42$ &
\\
 & & $m_s: -3.14$ & \\ \hline \hline
\multirow{2}{*}{C} &
\multirow{2}{*}{$\chi^2_\mathrm{masses} = 107.59$} &
$m_s: +1.03$ & \\
 & & $m_b: -10.32$ & \\*[1mm]
\hline
\end{tabular}
\caption{Description of the minimization results for the cases that fail.
The pull is defined as $H \left[ f_i \left( x \right) - \bar{O}_i \right]
\left. \left[ f_i \left( x \right) - \bar{O}_i \right] \right/ \delta_+ O_i
+ H \left[ \bar{O}_i - f_i \left( x \right) \right]
\left. \left[ f_i \left( x \right) - \bar{O}_i \right] \right/ \delta_- O_i$.}
\label{table-fails}
\end{table}

Only case D$_1$ is able to fit all the observables,
but all those good fits yield $\left| w_R / v_d \right| < 3 \times 10^{13}$.
This is unacceptable since,
with $\tan{\beta} = 10$,
$\left| v_d \right| = \left\langle H_d^0 \right\rangle_0
\approx 17.3\, \mathrm{GeV}$
then leads to $\left| w_R \right| \lesssim 5 \times 10^{14}\, \mathrm{GeV}$,
which is almost two orders of magnitude below the unification scale
$M_\mathrm{GUT} = 2 \times 10^{16}\, \mathrm{GeV}$.
If we enforce a more realistic $\left| w_R / v_d \right| > 10^{15}$
on case~D$_1$,
then we are only able to obtain poor fits with
$\chi^2_\mathrm{total} \gtrsim 60$.

\subsection{Case~A}

Case~A has much too many degrees of freedom,
so it is adequate to try and constrain it somewhat.
We follow ref.~\cite{GK2006},
in which real Yukawa-coupling matrices
(due to an additional $CP$ symmetry)
$F$,
$G$,
and $H$ were enforced and,
moreover,
$w_L = 0$ has been assumed,
thereby discarding the type-II seesaw mechanism.
Under these assumptions,
the authors of ref.~\cite{GK2006} have parameterized
\bs
\ba
M_d &=& \left( \begin{array}{ccc}
x + e^{i \zeta_d} a & e^{i \xi_d} f & e^{i \xi_d} g \\
- e^{i \xi_d} f & y + e^{i \zeta_d} b & e^{i \zeta_d} d \\
- e^{i \xi_d} g & e^{i \zeta_d} d & z + e^{i \zeta_d}  c
\end{array} \right), 
\\
M_\ell &=& \left( \begin{array}{ccc}
x - 3  e^{i \zeta_d} a & r_\ell e^{i \xi_\ell} f &
r_\ell e^{i \xi_\ell} g \\
- r_\ell e^{i \xi_\ell} f & y - 3  e^{i \zeta_d} b &
- 3  e^{i \zeta_d} d \\
- r_\ell e^{i \xi_\ell} g & - 3  e^{i \zeta_d} d &
z - 3  e^{i \zeta_d} c
\end{array} \right),
\\
M_u &=& \left( \begin{array}{ccc}
r_H x + r_F e^{i \zeta_u} a & r_u e^{i \xi_u} f & r_u e^{i \xi_u} g \\
- r_u e^{i \xi_u} f & r_H y + r_F e^{i \zeta_u} b &
r_F e^{i \zeta_u} d \\
- r_u e^{i \xi_u} g & r_F e^{i \zeta_u} d &
r_H z + r_F e^{i \zeta_u} c
\end{array} \right),
\\
M_D &=& \left( \begin{array}{ccc}
r_H x - 3 r_F e^{i \zeta_u} a & r_D e^{i \xi_D} f & r_D e^{i \xi_D} g \\
- r_D e^{i \xi_D} f & r_H y - 3 r_F e^{i \zeta_u} b &
- 3 r_F e^{i \zeta_u} d \\ 
- r_D e^{i \xi_D} g & - 3 r_F e^{i \zeta_u} d &
r_H z - 3 r_F e^{i \zeta_u} c
\end{array} \right),
\\
\left| \frac{w_R}{v_d} \right| \mnu &=&
\frac{1}{a \left( b c - d^2 \right)}\, M_D
\left( \begin{array}{ccc}
b c - d^2 & 0 & 0 \\
0 & a c & - a d \\
0 & - a d & a b
\end{array} \right)
M_D^T,
\label{pgiys}
\ea
\es
where
\bs
\ba
r_\ell e^{i \xi_\ell} &\equiv& \frac{\kappa_\ell}{\left| \kappa_d \right|},
\\
r_u e^{i \xi_u} &\equiv& \frac{\kappa_u}{\left| \kappa_d \right|},
\\
r_D e^{i \xi_D} &\equiv& \frac{\kappa_D}{\left| \kappa_d \right|}.
\\
r_H &\equiv& \left| \frac{k_u}{k_d} \right|,
\\
r_F e^{i \zeta_u} &\equiv& \frac{v_u}{\left| v_d \right|}.
\ea
\es
In this parameterization,
there are six phases
($\xi_\ell$, $\xi_u$, $\xi_D$, $\zeta_u$, $\xi_d$, and $\zeta_d$)
and 15 moduli
($x$, $y$, $z$, $a$, $b$, $c$,
$d$, $f$, $g$,
$r_\ell$, $r_H$, $r_u$, $r_F$, $r_D$, and $\left| w_R / v_d \right|$).

As usual,
we firstly fit the charged-fermion masses,
the mixing angles,
and $r^2_\mathrm{solar}$.
Secondly we adjust the factor $\left| w_R / v_d \right|$
in the left-hand side of equation~(\ref{pgiys}) in such a way that
$\left| m_3^2 - m_1^2 \right| = 2.5 \times 10^{-3}\, \mathrm{eV}^2$. 
Thirdly we compute $\left| v_d \right| = \left| v_d / w_R \right|
\left( 2 \times 10^{16}\, \mathrm{GeV} \right)$
and $\left| v_u \right| = r_F \left| v_d \right|$.
Finally,
we check that
\bs
\ba
\left| v_d \right|^2 &<& \left( \left\langle H_d^0 \right\rangle_0 \right)^2
\approx \left( 17.3\, \mathrm{GeV} \right)^2,
\\
\left| v_u \right|^2 &<& \left( \left\langle H_u^0 \right\rangle_0 \right)^2
\approx \left( 173\, \mathrm{GeV} \right)^2.
\ea
\es
We also check that
\be
\max \left( a^2, b^2, c^2, d^2 \right) < y^2 \left| v_d \right|^2
\ee
for some $1 < y < 10$;
we also require
\bs
\ba
\max \left( x^2, y^2, z^2 \right) + \left( 1 + \frac{r_\ell^2}{3} \right)
\max \left( f^2, g^2 \right) < y^2 \left[
\left( \left\langle H_d^0 \right\rangle_0 \right)^2
- \left| v_d \right|^2 \right],
\\
r_H^2 \max \left( x^2, y^2, z^2 \right) + \left( r_u^2 + \frac{r_D^2}{3} \right)
\max \left( f^2, g^2 \right) < y^2 \left[
\left( \left\langle H_u^0 \right\rangle_0 \right)^2
- \left| v_u \right|^2 \right].
\ea
\es

In ref.~\cite{GK2006} an explicit fit of case~A---under the above
restrictions $w_L = 0$ and real Yukawa-coupling matrices---to some data
was presented.
However,
the authors of ref.~\cite{GK2006} have used the charged-fermion masses
give in ref.~\cite{das} and have used the upper bound
on $\sin^2{\theta_{13}}$ that existed at the time.
We have attempted to fit case~A both to the updated charged-fermion masses
of ref.~\cite{zhou} and to the now extant value of $\sin^2{\theta_{13}}$.
We could achieve an excellent fit when the neutrino mass spectrum is normal
and a passable one when the mass spectrum is inverted;
those fits are presented in tables~\ref{tableAnormal} and~\ref{tableAinverted},
respectively.
%
\begin{table}[ht!]
\renewcommand{\arraystretch}{1.25}
\centering
\begin{tabular}{|c||c|}
\hline
parameter & value \\ \hline \hline
$x \left/ \, \mathrm{MeV} \right.$ & -0.476561625448 \\ \hline
$y \left/ \, \mathrm{MeV} \right.$ & -63.004302166872 \\ \hline
$z \left/ \, \mathrm{MeV} \right.$ & 410.084319821441 \\ \hline
$a \left/ \, \mathrm{MeV} \right.$ & -0.372645355981 \\ \hline
$b \left/ \, \mathrm{MeV} \right.$ & -91.581208223942 \\ \hline
$c \left/ \, \mathrm{MeV} \right.$ & -342.559232981562 \\ \hline
$d \left/ \, \mathrm{MeV} \right.$ & -172.410208448655 \\ \hline
$f \left/ \, \mathrm{MeV} \right.$ & -3.322328858814 \\ \hline
$g \left/ \, \mathrm{MeV} \right.$ & -0.261790479555 \\ \hline
$r_\ell e^{i \xi_\ell}$ &
$4.197350155392\, \exp{\left( i\, 3.160952468 \right)}$ \\ \hline
$r_u e^{i \xi_u}$ &
$6.938241672636\, \exp{\left( i\, 2.800761399333 \right)}$ \\ \hline
$r_D e^{i \xi_D}$ &
$5682.169770871835\, \exp{\left( i\, 4.151626745 \right)}$ \\ \hline
$r_F e^{i \zeta_u}$ &
$131.838888425156\, \exp{\left( i\, 3.114060300 \right)}$ \\ \hline
$r_H$ & 100.325400021876 \\ \hline
$\left. \xi_d \, \right/ \mathrm{rad}$ & 1.736825028772 \\ \hline
$\left. \zeta_d \, \right/ \mathrm{rad}$ & 2.935971894656 \\ \hline
$\left| w_R / v_d \right|$ & $1.90053716 \times 10^{16}$  \\ \hline
\end{tabular}
\caption{The values of the parameters for out best fit of case~A
with a normal neutrino mass spectrum.
The fit has $\chi^2_\mathrm{total} \approx 0.005$.}
\label{tableAnormal}
\end{table}
%
\begin{table}[ht!]
\renewcommand{\arraystretch}{1.25}
\centering
\begin{tabular}{|c||c|}
\hline
parameter & value \\ \hline \hline
$x \left/ \, \mathrm{MeV} \right.$ & 0.980345675289 \\ \hline
$y \left/ \, \mathrm{MeV} \right.$ & 13.317045360098 \\ \hline
$z \left/ \, \mathrm{MeV} \right.$ & 834.100031282343 \\ \hline
$a \left/ \, \mathrm{MeV} \right.$ & 1.230521136698 \\ \hline
$b \left/ \, \mathrm{MeV} \right.$ & 18.085613337982 \\ \hline
$c \left/ \, \mathrm{MeV} \right.$ & 82.58146771928 \\ \hline
$d \left/ \, \mathrm{MeV} \right.$ & 36.879698307221 \\ \hline
$f \left/ \, \mathrm{MeV} \right.$ & -2.572520483121 \\ \hline
$g \left/ \, \mathrm{MeV} \right.$ & 3.267046126672 \\ \hline
$r_\ell e^{i \xi_\ell}$ &
$5.389802407484\, \exp{\left( i\, 5.291743244393 \right)}$ \\ \hline
$r_u e^{i \xi_u}$ &
$9.506621363405\, \exp{\left( i\, 6.456544563269 \right)}$ \\ \hline
$r_D e^{i \xi_D}$ &
$19058.47748201563\, \exp{\left( i\, 4.698412501341 \right)}$ \\ \hline
$r_F e^{i \zeta_u}$ &
$93.384741884164\, \exp{\left( i\, 3.1289172068552067 \right)}$ \\ \hline
$r_H$ & 119.091394096965 \\ \hline
$\left. \xi_d \, \right/ \mathrm{rad}$ & 6.182757601569 \\ \hline
$\left. \zeta_d \, \right/ \mathrm{rad}$ & 4.027845889022 \\ \hline
$\left| w_R / v_d \right|$ & $9.0899658664 \times 10^{16}$  \\ \hline
\end{tabular}
\caption{The values of the parameters for out best fit of case~A
with an inverted neutrino mass spectrum.
This fit has $\chi^2_\mathrm{total} \approx 0.8$.}
\label{tableAinverted}
\end{table}
%

For the fit of table~\ref{tableAnormal} one has
$m_1 + m_2 + m_3 \approx 0.06\, \mathrm{eV}$. The fit of
table~\ref{tableAinverted} has $m_1 + m_2 + m_3 \approx 0.1\, \mathrm{eV}$.

\section{Conclusions}
\label{concl}

In this paper we have considered a supersymmetric $SO(10)$ GUT
in which the fermion masses are generated by renormalizable Yukawa couplings. 
Consequently,
the scalar multiplets under consideration belong to the irreps $\mathbf{10}$,
$\overline{\mathbf{126}}$,
and $\mathbf{120}$ of $SO(10)$.
We have assumed that there is a single scalar multiplet
belonging to each of these three irreps;
some further mild assumptions are listed in section~\ref{notation}.
We have analysed the prospects of imposing flavour symmetries in this scenario,
potentially making it predictive.
An exhaustive discussion has revealed 14 cases compatible with our scenario.
For the numerical examination of those cases we have used
the charged-fermion masses evaluated at the GUT scale
through renormalization-group running
in the context of the Minimal Supersymmetric Standard Model.
Interestingly,
the numerical analysis ruled out all 14 cases 
except case~A---see equation~(\ref{caseA})---and
case~B---see equation~(\ref{caseB}).
We have demonstrated that both cases~A and~B allow excellent fits to the data
when the neutrino mass spectrum is normal;
when that spectrum is inverted,
case~A can still fit the data but we were unable to find a fit for case~B.

Thus,
we have come to the conclusion that within the NMSGUT~\cite{nmsgut},
which has renormalizable Yukawa couplings just as the ones considered here, 
there are at most two possibilities
to reduce the number of Yukawa couplings through flavour symmetries,
while remaining in agreement with the data.

\begin{appendix}

\section{The scalar potential with an additional $\mathbf{45}$}
\label{potential}
\setcounter{equation}{0}
\renewcommand{\theequation}{A\arabic{equation}}

This appendix addresses two problems that may in general arise
in a renormalizable supersymmetric $SO(10)$ GUT
furnished with additional symmetries:
\begin{itemize}
\item How to promote the full mixing among the Higgs doublets
residing in the $\de$,
$\vi$,
and $\seb$ of $SO(10)$.
\item How to achieve the full breaking of $SO(10)$ to the SM gauge group
by using only renormalizable interactions. 
\end{itemize}
Both these problems can be solved in the NMSGUT,
but they are non-trivial in the context of our symmetry-furnished cases,
especially when the symmetry is larger than $\zz_2$.

\smallskip

According to refs.~\cite{Aulakh:2003kg,girdhar,Garg:2015mra},
in the NMSGUT there are five scalar irreps:
the $\de$,
the $\vi$,
the $\se$,
the $\seb$,
and the $\du$.
The $\de$,
the $\vi$,
and the $\seb$ have Yukawa couplings;
the $\se$ and the $\du$ do not.
The $\du$ is needed,
together with the $\se$ and $\seb$,
in order to break $SO(10)$ down to the SM gauge group.

In our models,
we propose to add to the NMSGUT one further scalar irrep---the $\qu$,
which is the adjoint of $SO(10)$.
The full superpotential is then\footnote{One may check that
no term is missing in equation~(\ref{wuiot})
by studying table~820 of ref.~\cite{naoki}.}
\ba
V_\mathrm{super} &=&
\lambda_1 \ \de \ \de
+ \lambda_2 \ \qu \ \qu
+ \lambda_3 \ \vi \ \vi
+ \lambda_4 \ \du \ \du
+ \lambda_5 \ \se \ \seb
\no & &
+ \lambda_6 \ \du \ \du \ \du
+ \lambda_7 \ \qu \ \qu \ \du
+ \lambda_8 \ \se \ \seb \ \du
\no & &
+ \lambda_9 \ \de \ \se \ \du 
+ \lambda_{10} \ \de \ \seb \ \du 
+ \lambda_{11} \ \vi \ \vi \ \du
+ \lambda_{12} \ \de \ \vi \ \du 
\no & &
+ \lambda_{13} \ \vi \ \se \ \du 
+ \lambda_{14} \ \vi \ \seb \ \du 
+ \lambda_{15} \ \se \ \seb \ \qu
\no & &
+ \lambda_{16} \ \de \ \vi \ \qu 
+ \lambda_{17} \ \vi \ \se \ \qu 
+ \lambda_{18} \ \vi \ \seb \ \qu.
\label{wuiot}
\ea
In order to go from the superpotential to the scalar potential
one must square the partial derivative relative to each superfield.
Thus,
the scalar potential is of the form
\ba
V &=&
\left| \lambda_1 \ \de
+ \lambda_9 \ \se \ \du
+ \lambda_{10} \ \seb \ \du
+ \lambda_{12} \ \vi \ \du
+ \lambda_{16} \ \vi \ \qu \right|^2
\no & &
+ \left| \lambda_2 \ \qu
+ \lambda_7 \ \qu \ \du
+ \lambda_{15} \ \se \ \seb
\right. \no & & \left.
+ \lambda_{16} \ \de \ \vi
+ \lambda_{17} \ \vi \ \se
+ \lambda_{18} \ \vi \ \seb \right|^2
\no & &
+ \left| \lambda_3 \ \vi
+ \lambda_{11} \ \vi \ \du
+ \lambda_{12} \ \de \ \du
+ \lambda_{13} \ \se \ \du
+ \lambda_{14} \ \seb \ \du
\right. \no & & \left.
+ \lambda_{16} \ \de \ \qu
+ \lambda_{17} \ \se \ \qu
+ \lambda_{18} \ \seb \ \qu \right|^2
\no & &
+ \left| \lambda_4 \ \du
+ \lambda_6 \ \du \ \du
+ \lambda_7 \ \qu \ \qu
+ \lambda_8 \ \se \ \seb
+ \lambda_9 \ \de \ \se
\right. \no & & \left.
+ \lambda_{10} \ \de \ \seb
+ \lambda_{11} \ \vi \ \vi
+ \lambda_{12} \ \de \ \vi
+ \lambda_{13} \ \vi \ \se
+ \lambda_{14} \ \vi \ \seb \right|^2
\no & &
+ \left| \lambda_5 \ \seb
+ \lambda_8 \ \seb \ \du
+ \lambda_9 \ \de \ \du
+ \lambda_{13} \ \vi \ \du
\right. \no & & \left.
+ \lambda_{15} \ \seb \ \qu
+ \lambda_{17} \ \vi \ \qu
\right|^2
\no & &
+ \left| \lambda_5 \ \se
+ \lambda_8 \ \se \ \du
+ \lambda_{10} \ \de \ \du
+ \lambda_{14} \ \vi \ \du
\right. \no & & \left.
+ \lambda_{15} \ \se \ \qu
+ \lambda_{18} \ \vi \ \qu
\right|^2.
\label{pot}
\ea
The $\de$ and the $\vi$ do not have any component
which is invariant under the SM gauge group,
therefore they are not allowed to acquire a VEV at the GUT scale.
Thus,
at the GUT scale the relevant potential is just
\ba
V_\mathrm{GUT} &=&
\left| \lambda_9 \ \se \ \du
+ \lambda_{10} \ \seb \ \du \right|^2
+ \left| \lambda_2 \ \qu
+ \lambda_7 \ \qu \ \du
+ \lambda_{15} \ \se \ \seb
\right|^2
\no & &
+ \left| \lambda_{13} \ \se \ \du
+ \lambda_{14} \ \seb \ \du
+ \lambda_{17} \ \se \ \qu
+ \lambda_{18} \ \seb \ \qu \right|^2
\no & &
+ \left| \lambda_4 \ \du
+ \lambda_6 \ \du \ \du
+ \lambda_7 \ \qu \ \qu
+ \lambda_8 \ \se \ \seb
\right|^2
\no & &
+ \left| \lambda_5 \ \seb
+ \lambda_8 \ \seb \ \du
+ \lambda_{15} \ \seb \ \qu
\right|^2
\no & &
+ \left| \lambda_5 \ \se
+ \lambda_8 \ \se \ \du
+ \lambda_{15} \ \se \ \qu
\right|^2.
\label{pgut}
\ea

In our case~A there is a symmetry $\vi \to - \vi$.
We must extend it and make $\qu \to - \qu$ too.
Then the symmetry implies
$\lambda_{12} = \lambda_{13} = \lambda_{14} = \lambda_{15} = 0$.
Equation~(\ref{pot}) becomes
\ba
V^\mathrm{(case\, A)} &=&
\left| \lambda_1 \ \de
+ \lambda_9 \ \se \ \du
+ \lambda_{10} \ \seb \ \du
+ \lambda_{16} \ \vi \ \qu \right|^2
\no & &
+ \left| \lambda_2 \ \qu
+ \lambda_7 \ \qu \ \du
+ \lambda_{16} \ \de \ \vi
+ \lambda_{17} \ \vi \ \se
+ \lambda_{18} \ \vi \ \seb \right|^2
\no & &
+ \left| \lambda_3 \ \vi
+ \lambda_{11} \ \vi \ \du
+ \lambda_{16} \ \de \ \qu
+ \lambda_{17} \ \se \ \qu
+ \lambda_{18} \ \seb \ \qu \right|^2
\no & &
+ \left| \lambda_4 \ \du
+ \lambda_6 \ \du \ \du
+ \lambda_7 \ \qu \ \qu
+ \lambda_8 \ \se \ \seb
+ \lambda_9 \ \de \ \se
\right. \no & & \left.
+ \lambda_{10} \ \de \ \seb
+ \lambda_{11} \ \vi \ \vi
\right|^2
\no & &
+ \left| \lambda_5 \ \seb
+ \lambda_8 \ \seb \ \du
+ \lambda_9 \ \de \ \du
+ \lambda_{17} \ \vi \ \qu
\right|^2
\no & &
+ \left| \lambda_5 \ \se
+ \lambda_8 \ \se \ \du
+ \lambda_{10} \ \de \ \du
+ \lambda_{18} \ \vi \ \qu
\right|^2
\label{buipy}
\ea
and equation~(\ref{pgut}) becomes
\ba
V_\mathrm{GUT}^\mathrm{(case\, A)} &=&
\left| \lambda_9 \ \se \ \du
+ \lambda_{10} \ \seb \ \du \right|^2
+ \left| \lambda_2 \ \qu
+ \lambda_7 \ \qu \ \du
\right|^2
\no & &
+ \left| \lambda_{17} \ \se \ \qu
+ \lambda_{18} \ \seb \ \qu \right|^2
\no & &
+ \left| \lambda_4 \ \du
+ \lambda_6 \ \du \ \du
+ \lambda_7 \ \qu \ \qu
+ \lambda_8 \ \se \ \seb
\right|^2
\no & &
+ \left| \lambda_5 \ \seb
+ \lambda_8 \ \seb \ \du
\right|^2
+ \left| \lambda_5 \ \se
+ \lambda_8 \ \se \ \du
\right|^2.
\label{giuto}
\ea
The third line of equation~(\ref{buipy}) indicates that
the $\vi$ fully mixes with both the $\de$ and the $\seb$.
The first and last lines of equation~(\ref{giuto}) indicate that
the potential for the $\qu$,
$\se$,
$\seb$,
and $\du$ allows all of them to acquire VEVs.

In our case~B there is a symmetry $\de \to - \de, \ \vi \to - \vi, \
\qu \to - \qu$.
This implies $\lambda_9 = \lambda_{10}
= \lambda_{13} = \lambda_{14} = \lambda_{15} = \lambda_{16} = 0$.
Equation~(\ref{pot}) becomes
\ba
V^\mathrm{(case\, B)} &=&
\left| \lambda_1 \ \de + \lambda_{12} \ \vi \ \du \right|^2
\no & &
+ \left| \lambda_2 \ \qu
+ \lambda_7 \ \qu \ \du
+ \lambda_{17} \ \vi \ \se
+ \lambda_{18} \ \vi \ \seb \right|^2
\no & &
+ \left| \lambda_3 \ \vi
+ \lambda_{11} \ \vi \ \du
+ \lambda_{12} \ \de \ \du
+ \lambda_{17} \ \se \ \qu
+ \lambda_{18} \ \seb \ \qu \right|^2
\no & &
+ \left| \lambda_4 \ \du
+ \lambda_6 \ \du \ \du
+ \lambda_7 \ \qu \ \qu
\right. \no & & \left.
+ \lambda_8 \ \se \ \seb
+ \lambda_{11} \ \vi \ \vi
+ \lambda_{12} \ \de \ \vi
\right|^2
\no & &
+ \left| \lambda_5 \ \seb
+ \lambda_8 \ \seb \ \du
+ \lambda_{17} \ \vi \ \qu
\right|^2
\no & &
+ \left| \lambda_5 \ \se
+ \lambda_8 \ \se \ \du
+ \lambda_{18} \ \vi \ \qu
\right|^2
\label{buify}
\ea
and equation~(\ref{pgut}) becomes
\ba
V_\mathrm{GUT}^\mathrm{(case\, B)} &=&
\left| \lambda_2 \ \qu
+ \lambda_7 \ \qu \ \du
\right|^2
+ \left| \lambda_{17} \ \se \ \qu
+ \lambda_{18} \ \seb \ \qu \right|^2
\no & &
+ \left| \lambda_4 \ \du
+ \lambda_6 \ \du \ \du
+ \lambda_7 \ \qu \ \qu
+ \lambda_8 \ \se \ \seb
\right|^2
\no & &
+ \left| \lambda_5 \ \seb
+ \lambda_8 \ \seb \ \du
\right|^2
+ \left| \lambda_5 \ \se
+ \lambda_8 \ \se \ \du
\right|^2.
\label{gsuto}
\ea
The third line of equation~(\ref{buify}) indicates that
the $\vi$ fully mixes with both the $\de$ and the $\seb$.
Equation~(\ref{gsuto}) demonstrates that the potential for the $\qu$,
$\se$,
$\seb$,
and $\du$ allows all of them to acquire VEVs.

Case~C may be treated in a similar fashion.
Our remaining cases have symmetries $\zz_n$ with $n > 2$
and are much more problematic.
Anyway,
we do not need to worry about those cases since we already know
that they are unable to fit the phenomenological data.

\section{Investigation of a second symmetry}
\label{investigation}
\setcounter{equation}{0}
\renewcommand{\theequation}{B\arabic{equation}}

In this appendix we take all 13 cases of subsection~\ref{single} and consider,
for each of them,
the possibility of a second flavour symmetry defined in equation~(\ref{S1}).
Without loss of generality we set $e^{i\beta_1} = 1$ in that equation.

The conclusion of this appendix is that,
beyond those 13 cases,
only one new case arises which does not contradict our assumptions---case~E
in equation~(\ref{A3}).

\subsection{Cases A$_1$, A$^\prime_1$, A$^{\prime\prime}_1$, and A$_2$}

In all these four cases,
\be
\label{ggg}
G =
\left( \begin{array}{ccc}
 0 & 0 & d \\
 0 & 0 & 0 \\
- d & 0 & 0
\end{array} \right)
\ee
with $d \neq 0$.
Since
\be
\label{G1}
X^T G X = G
\ \Leftrightarrow \
G X = X^* G,
\ee
we find 
\be
\label{G2}
x_{12} = x_{21} = x_{23} = x_{32} = 0,
\quad
X = \left( \begin{array}{ccc}
x_{11} & 0 & x_{13} \\ 0 & x_{22} & 0 \\ - x_{13}^\ast & 0 & x_{11}^\ast
\end{array} \right).
\ee

In these four cases the matrix $F$ has the form  
\be
\label{F1}
F = F_1 \equiv \left( \begin{array}{ccc}
0 & a & 0 \\
a & 0 & 0 \\
0 & 0 & b
\end{array} \right). 
\ee
We require $\det F \neq 0$,
hence $a \neq 0$ and $b \neq 0$.
Using
\be
\label{ugipt}
e^{i \gamma_1} F X = X^\ast F
\ee
with the matrix $X$ of equation~(\ref{G2}),
we obtain that $X$ must be diagonal:
\be
X = \mbox{diag} \left( e^{i \gamma_1 / 2},\ e^{- 3 i \gamma_1 / 2},\ e^{- i \gamma_1 / 2}
\right).
\ee

Now we look for the consequences of
\be
\label{bvjoty}
e^{i \alpha_1} H X = X^\ast H.
\ee
With a diagonal $X$,
equation~(\ref{bvjoty}) can only force
either one or more matrix elements of $H$ to be zero.
In the case A$_1$,
if one sets one matrix element of $H$ to zero
then one simply recovers the cases A$^\prime_1$ and A$^{\prime\prime}_1$.
In the cases A$^\prime_1$ and A$^{\prime\prime}_1$,
the number of non-vanishing elements of $H$ is already minimal.
In the case A$_2$ we have $X^T H X = e^{- i \gamma_1} H$,
therefore either $\alpha_1 = \gamma_1$
and $H$ is not restricted by $\mathcal{S}_1$
or $\alpha_1 \neq \gamma_1$ and $H=0$,
which is excluded by our assumptions.

In summary,
departing from cases A$_1$,
A$^\prime_1$,
A$^{\prime\prime}_1$,
or A$_2$ no new cases can ensue from a second symmetry.

\subsection{Cases D$_2$, D$_3$, D$^\prime_2$, and D$^\prime_3$}

In these cases equation~(\ref{ggg}) is still valid,
therefore equation~(\ref{G2}) also holds.
In all four cases
\be
\label{F2}
F = F_2 \equiv \left( \begin{array}{ccc}
b & 0 & 0 \\
0 & 0 & a \\
0 & a & 0
\end{array} \right),
\ee
with $a \neq 0$ and $b \neq 0$.
Using equation~(\ref{ugipt}) then yields
\be
X = \mbox{diag} \left( e^{- i \gamma_1 / 2},\ e^{- 3 i \gamma_1 / 2},\ e^{i \gamma_1 / 2}
\right),
\ee
\textit{i.e.}~$X$ is once again diagonal.

We next consider equation~(\ref{bvjoty}).
In case D$_3$ we obtain
\be
H =
\left( \begin{array}{ccc}
0 & 0 & r \\ 0 & s & 0 \\ r & 0 & 0
\end{array} \right) =
e^{i \alpha_1} X^T H X =
e^{i \alpha_1} \left( \begin{array}{ccc}
0 & 0 & r \\ 0 & e^{- 3 i \gamma_1} s & 0 \\ r & 0 & 0
\end{array} \right).
\ee
In case D$_2$ we have
\be
H =
\left( \begin{array}{ccc}
0 & r & 0 \\ r & 0 & 0 \\ 0 & 0 & s
\end{array} \right) =
e^{i \alpha_1} X^T H X =
e^{i \alpha_1} \left( \begin{array}{ccc}
0 & e^{- 2 i \gamma_1} r & 0 \\ e^{- 2 i \gamma_1} r & 0 & 0 \\ 0 & 0 & e^{i \gamma_1} s
\end{array} \right).
\ee
Thus,
equation~(\ref{bvjoty}) can at most set either $r=0$ or $s=0$.
If $s=0$ then one recovers case D$^\prime_2$ from case D$_2$
and case D$^\prime_3$ from case D$_3$.
If $r=0$ then,
through an interchange of the first and third generations,
one recovers case A$^\prime_1$ from case D$_2$
and case A$^{\prime\prime}_1$ from case D$_3$.
Therefore,
no new cases arise from the enforcement of the symmetry $\mathcal{S}_1$
on any of these four cases.

\subsection{Cases D$_1$ and D$^\prime_1$}

Equations~(\ref{ggg}) and~(\ref{G2}) once again hold.
Now
\be
\label{F3}
F = F_3 \equiv \left( \begin{array}{ccc}
0 & 0 & a \\
0 & b & 0 \\
a & 0 & 0
\end{array} \right),
\ee
with $a \neq 0$ and $b \neq 0$.
Equation~(\ref{ugipt}) then yields that either $e^{i \gamma_1} = +1$ and
\be
\label{x1}
X = \mbox{diag} \left( e^{i \psi},\ \pm 1,\ e^{- i \psi} \right)
\ee
or $e^{i \gamma_1} = -1$ and
\be
\label{x2}
X = \left( \begin{array}{ccc}
0 & 0 & e^{i \varphi} \\ 0 & \pm i & 0 \\ - e^{- i \varphi} & 0 & 0
\end{array} \right).
\ee
In case D$_1$ and with equation~(\ref{x1}) one obtains
\be
H = \left( \begin{array}{ccc}
0 & r & 0 \\ r & 0 & 0 \\ 0 & 0 & s
\end{array} \right)
= e^{i \alpha_1} X^T H X = e^{i \alpha_1} \left( \begin{array}{ccc}
0 & \pm e^{i \psi} r & 0 \\ \pm e^{i \psi} r & 0 & 0 \\ 0 & 0 & e^{- 2 i \psi} s
\end{array} \right).
\label{dsjipt}
\ee
With equation~(\ref{x2}) one arrives instead at
\be
H = \left( \begin{array}{ccc}
0 & r & 0 \\ r & 0 & 0 \\ 0 & 0 & s
\end{array} \right)
= e^{i \alpha_1} X^T H X = e^{i \alpha_1}
\left( \begin{array}{ccc}
e^{- 2 i \varphi} s & 0 & 0 \\
0 & 0 & \pm i e^{i \varphi} r \\
0 & \pm i e^{i \varphi} r & 0
\end{array} \right).
\ee
Thus,
the possibility~(\ref{x2}) implies $H = 0$,
which contradicts our assumptions.
With equation~(\ref{dsjipt}) then either $s=0$ and one recovers case D$^\prime_1$
or $r=0$ and the second generation decouples.
We conclude that the enforcement of the symmetry $\mathcal{S}_1$
on cases D$_1$ and D$^\prime_1$ cannot lead to new cases.

\subsection{Cases A and B, step 1: $X$ may be chosen to be diagonal}

In cases A and B we may perform a weak-basis transformation
such that $G$ acquires the form~(\ref{ggg})
while the forms of $H$ and $F$ are kept unchanged:
\bs
\label{sgyit}
\ba
\mbox{case~A}: & &
H \sim \left( \begin{array}{ccc}
\times & \times & 0 \\
\times & \times & 0 \\
0 & 0 & \times
\end{array} \right),
\quad
G \sim \left( \begin{array}{ccc}
0 & 0 & \times \\
0 & 0 & 0 \\
\times & 0 & 0
\end{array} \right),
\quad
F \sim \left( \begin{array}{ccc}
\times & \times & 0 \\
\times & \times & 0 \\
0 & 0 & \times
\end{array} \right); \hspace*{11mm}
\\
\mbox{case B}: & &
H \sim \left( \begin{array}{ccc}
0 & 0 & \times \\
0 & 0 & \times \\
\times & \times & 0
\end{array} \right),
\quad
G \sim \left( \begin{array}{ccc}
0 & 0 & \times \\
0 & 0 & 0 \\
\times & 0 & 0
\end{array} \right),
\quad
F \sim \left( \begin{array}{ccc}
\times & \times & 0 \\
\times & \times & 0 \\
0 & 0 & \times
\end{array} \right).
\ea
\es
This is achieved through a unitary rotation of the first and second generations,
which does not alter the matrix $W = \mbox{diag} \left( +1,\ +1,\ -1 \right)$
for these cases.
In the new basis~(\ref{sgyit}),
equation~(\ref{G2}) holds.

Next we consider equation~(\ref{ugipt}).
With $F \equiv \left( f_{ij} \right)$,
it reads
\be
\label{fx}
e^{i \gamma_1} \left( \begin{array}{ccc}
f_{11} x_{11} & f_{12} x_{22} & f_{11} x_{13} \\
f_{12} x_{11} & f_{22} x_{22} & f_{12} x_{13} \\
- f_{33} x_{13}^\ast & 0 & f_{33} x_{11}^\ast 
\end{array} \right) =
\left( \begin{array}{ccc}
f_{11} x_{11}^* & f_{12} x_{11}^* & f_{33} x_{13}^* \\
f_{12} x_{22}^* & f_{22} x_{22}^* & 0              \\
- f_{11} x_{13} & - f_{12} x_{13} & f_{33} x_{11} 
\end{array} \right).
\ee
Let us suppose that $X$ is not diagonal,
\textit{i.e.}~that $x_{13}$ is nonzero.
Then equation~(\ref{fx}) tells us that $f_{12} = 0$,
\textit{i.e.}~that $F$ is diagonal.
Now we invoke $e^{i\alpha_1} H X = X^* H$.
In case~A the matrix $H \equiv \left( h_{ij} \right)$
has the same form as the matrix $F$,
hence we may conclude,
from the analogue of equation~(\ref{fx}),
that $h_{12} = 0$ just as $f_{12} = 0$,
\textit{i.e.}~$H$ is diagonal too.
But then the second generation decouples,
which runs against our assumptions.
For case~B the equation $e^{i\alpha_1} H X = X^* H$ reads
\be
e^{i\alpha_1}
\left( \begin{array}{ccc}
- h_{13} x_{13}^\ast & 0 & h_{13} x_{11}^\ast \\
- h_{23} x_{13}^\ast & 0 & h_{23} x_{11}^\ast \\
h_{13} x_{11} & h_{23} x_{22} & h_{13} x_{13} 
\end{array} \right) = \left( \begin{array}{ccc}
h_{13} x_{13}^* & h_{23} x_{13}^* & h_{13} x_{11}^* \\
0 & 0 & h_{23} x_{22}^* \\
h_{13} x_{11} & h_{23} x_{11} & - h_{13} x_{13} 
\end{array} \right),
\ee
hence $h_{23} = 0$ and the second generation decouples.

We conclude that the hypothesis $x_{13} \neq 0$
leads to a contradiction with our assumptions.
Thus,
cases~A and~B do not admit a non-diagonal $X$.

\subsection{Cases A and B, step 2: the forms of $F$ and $X$}

With a diagonal matrix $X$,
the equation $X^T F X e^{i \gamma_1} = F$ yields
\bs
\ba
x_{11}^2 f_{11} e^{i \gamma_1} &=& f_{11}, \\
x_{22}^2 f_{22} e^{i \gamma_1} &=& f_{22}, \\
x_{11} x_{22} f_{12} e^{i \gamma_1} &=& f_{12}, \\
x_{33}^2 f_{33} e^{i \gamma_1} &=& f_{33}. \label{bipy}
\ea
\es
Since $\det{F} \neq 0$,
$f_{33}$ cannot vanish.
Therefore,
equation~(\ref{bipy}) gives $x_{33} = \varepsilon e^{- i \gamma_1 / 2}$,
where $\varepsilon = \pm 1$.

Let us firstly suppose that $x_{11} = x_{22}$.
In this case we must have $x_{11}^2 = e^{- i \gamma_1}$,
else $f_{11} = f_{22} = f_{12} = 0$ and $\det{F} = 0$.
Consequently,
$x_{11} = \eta e^{- i \gamma_1 / 2}$,
where $\eta = \pm 1$.
In this case the matrix $F$ cannot be restricted any further by $\mathcal{S}_1$.

Since
$X = e^{- i \gamma_1 / 2}\ \mbox{diag} \left( \eta,\ \eta,\ \varepsilon \right)$,
$X^T H X = e^{- i \gamma_1} H$ in case~A
and $X^T H X = \varepsilon \eta e^{- i \gamma_1} H$ in case~B.
This means that 
the equation $e^{i \alpha_1} X^T H X = H$
either does not restrict $H$ any further,
or it enforces $H=0$
(depending on the choice for $e^{i\alpha_1}$).
Since $H=0$ runs against our assumptions,
we conclude that,
with $x_{11} = x_{22}$,
the symmetry $\mathcal{S}_1$ does not restrict the Yukawa-coupling matrices
any further,
\textit{i.e.}~it does not lead to any new cases.

So we are lead to consider $x_{11} \neq x_{22}$.
Then,
only two possibilities for $X$ remain,
which are compatible with $\det{F} \neq 0$:
either
\bs
\label{a}
\ba
X = X_a &\equiv& e^{- i \gamma_1 / 2}\
\diag \left( \eta,\ -\eta,\ \varepsilon \right), \\
F = F_a &\equiv& \diag \left( f_{11},\ f_{22},\ f_{33} \right),
\ea
\es
or
\bs
\label{b}
\ba
X = X_b &\equiv& e^{- i \gamma_1 / 2}\
\diag \left( e^{i\rho},\ e^{-i\rho},\ \varepsilon \right), \\
F &=& F_1,
\ea
\es
with $F_1$ given by equation~(\ref{F1}) and $e^{2 i \rho} \neq 1$.

We must remember that $X$ must be of the form~(\ref{G2}),
\textit{viz.}~that $x_{33} = x_{11}^\ast$.
Therefore,
\bs
\ba
e^{i \gamma_1} = e^{- i \gamma_1} = \eta \varepsilon
& & \mbox{if}\ X = X_a,
\\
e^{i \left( \gamma_1 - \rho \right)} = e^{i \left( \rho - \gamma_1 \right)} =
\varepsilon 
& & \mbox{if}\ X = X_b.
\ea
\es

\subsection{Cases A and B, step 3: the form of $H$}

\paragraph{Case A, $X = X_a$, $F = F_a$:}
In this case the equation $e^{i \alpha_1} X^T H X = H$ gives
\be
e^{i \left( \alpha_1 - \gamma_1 \right)}
\left( \begin{array}{ccc}
h_{11}  & -h_{12} & 0 \\
-h_{12}  & h_{22} & 0 \\
0 & 0 & h_{33}
\end{array} \right)
=
\left( \begin{array}{ccc}
h_{11}  & h_{12} & 0 \\
h_{12}  & h_{22} & 0 \\
0 & 0 & h_{33}
\end{array} \right).
\ee
If $e^{i \left( \alpha_1 - \gamma_1 \right)} \neq \pm 1$,
then $H = 0$ contradicts our assumptions.
If $e^{i \left( \alpha_1 - \gamma_1 \right)} = 1$,
then $h_{12} = 0$ and the second generation decouples.
If $e^{i \left( \alpha_1 - \gamma_1 \right)} = -1$,
then we get case E---see equations~(\ref{A3}),
because the choice $e^{i \gamma_1} = e^{i \gamma_1/2} = \eta = \varepsilon = 1$
indeed leads to the symmetry $\zz_2^{(2)}$ of that equation.

\paragraph{Case B, $X = X_a$, $F = F_a$:}
In this case the equation $e^{i \alpha_1} X^T H X = H$ gives
\be
e^{i \left( \alpha_1 - \gamma_1 \right)} \eta \varepsilon
\left( \begin{array}{ccc}
0  & 0 & h_{13} \\
0  & 0 & -h_{23} \\
h_{13} & -h_{23} & 0
\end{array} \right)
=
\left( \begin{array}{ccc}
0  & 0 & h_{13} \\
0  & 0 & h_{23} \\
h_{13} & h_{23} & 0
\end{array} \right).
\ee
In order to avoid decoupling of the second generation,
we must choose
$e^{i \left( \alpha_1 - \gamma_1 \right)} \eta \varepsilon = -1$ and 
$h_{13} = 0$.
We then obtain a case which is equivalent to case~E
after the interchange of the first and third generations.

\paragraph{Case A, $X = X_b$, $F = F_1$:}
In this case the equation $e^{i \alpha_1} X^T H X = H$ gives
\be
e^{i \left( \alpha_1 - \gamma_1 \right)}
\left( \begin{array}{ccc}
e^{2 i \rho} h_{11} & h_{12} & 0 \\
h_{12} & e^{- 2 i \rho} h_{22} & 0 \\
0 & 0 & h_{33}
\end{array} \right)
=
\left( \begin{array}{ccc}
h_{11} & h_{12} & 0 \\
h_{12} & h_{22} & 0 \\
0 & 0 & h_{33}
\end{array} \right).
\ee
Since $e^{2 i \rho} \neq 1$,
through a choice of the phases
we may achieve either case~A$_1$ or case A$^\prime_1$
or case A$^{\prime\prime}_1$ or case A$_2$;
no new case arises.

\paragraph{Case B, $X = X_b$, $F = F_1$:}
In this case the equation $e^{i \alpha_1} X^T H X = H$ gives
\be
e^{i \left( \alpha_1 - \gamma_1 \right)}\, \varepsilon
\left( \begin{array}{ccc}
0 & 0 & e^{i \rho} h_{13} \\
0 & 0 & e^{- i \rho} h_{23} \\
e^{i \rho} h_{13} & e^{- i \rho} h_{23} & 0
\end{array} \right) =
\left( \begin{array}{ccc}
0 & 0 & h_{13} \\
0 & 0 & h_{23} \\
h_{13} & h_{23} & 0
\end{array} \right).
\ee
In order to avoid decoupling of the second generation
we must choose $e^{i(\alpha_1-\gamma_1-\rho)}\, \varepsilon = 1$ and $h_{13} = 0$;
this case is equivalent to D$^\prime_2$
through the interchange of the first and third generations.

\subsection{Case C}

In case~C,
it is convenient to choose a weak basis where
\be
\label{ytuire}
H \sim \left( \begin{array}{ccc}
0 & \times & 0 \\
\times & 0 & 0 \\
0 & 0 & 0
\end{array} \right),
\quad
G \sim \left( \begin{array}{ccc}
0 & 0 & \times \\
0 & 0 & 0 \\
\times & 0 & 0
\end{array} \right),
\quad
F \sim \left( \begin{array}{ccc}
\times & 0 & \times \\
0 & \times & 0 \\
\times & 0 & \times
\end{array} \right).
\ee
This weak basis is achieved,
starting from the form~(\ref{HFG-C}) of the matrices $H$,
$G$,
and $F$,
through a unitary rotation mixing the first and third generations;
such a rotation does not alter the matrix $W$ in equation~(\ref{S0-C}).

With $G$ of equation~(\ref{ytuire}) we know that $X$ has to
obey equation~(\ref{G2}). 
It is then easy to see that 
$H X e^{i \alpha_1} = X^\ast H$ requires $X$ to be diagonal with 
$e^{i\alpha_1} x_{11} = x_{22}^*$. Therefore, $X$ can be parameterized as
\be
X = \mbox{diag} \left( e^{i \psi},\ e^{- i \left( \alpha_1 + \psi \right)},\
e^{- i \psi} \right).
\ee
With this $X$,
the equation $X^T F X e^{i \gamma_1} = F$ can only force
one or more matrix elements of $F$ to be zero.

If $e^{i\gamma_1} \neq 1$, one obtains $f_{13} = f_{31} = 0$ and, therefore,
case E. Then, because of $\det{F} \neq 0$, 
all $f_{ii}$ must be non-zero and it is easy to show 
that this leads to $e^{i\gamma_1} = -1$, $e^{i\psi} = i \eta$, and 
$e^{i\alpha_1} = \varepsilon$  
with $\eta^2 = \varepsilon^2 = 1$.
Summarizing, we have 
\be
X = i\eta \left( 1, -\varepsilon, -1 \right), \quad
e^{i\alpha_1} = \varepsilon, \quad
e^{i\beta_1} = +1, \quad
e^{i\gamma_1} = -1.
\ee
By choosing $\varepsilon = -1$ and absorbing $(i\eta)^2 = -1$ into the phase
factors, we arrive at $\zz_2^{(1)}$ of equation~(\ref{A3}). Note that in this
subsection the symmetry $\mathcal{S}_0$ is given by $\zz_2^{(2)}$ and 
$\mathcal{S}_1$ by $\zz_2^{(1)}$, since we started from case~C. Thus, in the
present subsection, the notation for $\zz_2^{(1)}$ and $\zz_2^{(2)}$ 
is exchanged compared to equation~(\ref{A3}).

Moving to $e^{i\gamma_1} = 1$ and taking again into account $\det F \neq 0$, 
we have $f_{22} \neq 0$. However,  
it is neither possible to enforce $f_{11} = 0$ while keeping $f_{33} \neq 0$
nor to enforce $f_{33} = 0$ while keeping $f_{11} \neq 0$;
with $f_{11} = f_{33} = 0$ one recovers case D$^\prime_1$.

One thus concludes that enforcing an extra symmetry on case~C
can only lead to 
cases~E or~$D^\prime_1$,
or else to a violation of our assumptions.

\section{Precise definition of the matrices $F$, $G$, $H$}
\label{inequalities}
\setcounter{equation}{0}
\renewcommand{\theequation}{C\arabic{equation}}

The aim of this appendix is to precisely define the matrices $F$,
$G$,
and $H$ through equations~\eqref{hhggff} and thereby to extract
the useful inequalities~\eqref{vhpty},
which we employ in subsection~\ref{ineq}.

The MSSM contains two Higgs doublets,
$H_d$ and $H_u$,
with hypercharges $+1/2$ and $-1/2$,
respectively.
Their corresponding VEVs are $v \cos{\beta}$ and $v \sin{\beta}$,
respectively,
where $v = 174\, \mathrm{ GeV}$. 
When one neglects the effects of the electroweak scale,
these two doublets are,
by assumption,
the only scalar zero-modes extant at the GUT scale;
this requires a minimal finetuning condition~\cite{AG02,girdhar}. 
Each of the scalar irreps $\mathbf{10}$,
$\overline{\mathbf{126}}$,
$\mathbf{126}$,
and $\mathbf{210}$
contains one doublet with the quantum numbers of $H_d$;
the $\mathbf{120}$ contains two such doublets.
The doublet $H_d$ is a superposition of these six doublets
with amplitudes $\bar\alpha_j$ ($j = 1,\ldots,6$). 
Let $\alpha_j$ denote the analogous coefficients for $H_u$.
The normalization conditions are
\be
\label{normalization}
\sum_{j=1}^6 \left| \bar\alpha_j \right|^2 =
\sum_{j=1}^6 \left| \alpha_j \right|^2 = 1.
\ee
It follows from equations~\eqref{normalization} that
\bs
\label{alpha}
\ba
\left| \bar\alpha_1 \right|^2 + \left| \bar\alpha_2 \right|^2
+ \left| \bar\alpha_5 \right|^2  + \left| \bar\alpha_6 \right|^2 &\leq& 1,
\\
\left| \alpha_1 \right|^2 + \left| \alpha_2 \right|^2
+ \left| \alpha_5 \right|^2 + \left| \alpha_6 \right|^2 &\leq& 1.
\ea
\es
The inequalities~\eqref{alpha} only involve the amplitudes
of the doublets contained in the $\de$,
$\seb$,
and $\vi$.

Taking into account that the $\mathbf{126}$ and the $\mathbf{210}$
have no Yukawa couplings,
the Dirac mass matrices are given by 
\bs
\label{M}
\ba
M_a &=&
v \cos{\beta} \left[ 
c^a_1 \bar\alpha_1 Y_{10} + c^a_2 \bar\alpha_2 Y_{\overline{126}} +
\left( c^a_5 \bar\alpha_5 + c^a_6 \bar\alpha_6 \right) Y_{120} \right]
\quad (a = d,\, \ell),
\label{Ma} \\*[1mm]
M_b &=&
v \sin{\beta} \left[ 
c^b_1 \alpha_1 Y_{10} + c^b_2 \alpha_2 Y_{\overline{126}} +
\left( c^b_5 \alpha_5 + c^b_6 \alpha_6 \right) Y_{120} \right]
\quad (b = u,\, D),
\label{Mb}
\ea
\es
with Yukawa-coupling matrices $Y_{10}$,
$Y_{\overline{126}}$,
and $Y_{120}$ and Clebsch--Gordan coefficients $c^{a,b}_j$;
the latter derive from the $SO(10)$-invariant
Yukawa couplings~\cite{AG041,AG042}.
The absolute values of the Clebsch--Gordan coefficients
have no physical meaning
and some of their phases are convention dependent.
With our conventions,\footnote{See the appendix of ref.~\cite{GK2006}.}
the required information reads
\bs
\label{c}
\ba
& & c^d_1 = c^u_1 = c^\ell_1 = c^D_1,
\\
& & 3 c^d_2 = - 3 c^u_2 = - c^\ell_2 = c^D_2,
\\*[1mm]
& & \sqrt{3}\, c^d_5 = - \sqrt{3}\, c^u_5 = \sqrt{3}\, c^\ell_5
= - \sqrt{3}\, c^D_5 = 3 c^d_6 = 3 c^u_6 = - c^\ell_6 = - c^D_6.
\ea
\es

In order to make contact with the mass formulas,
we define
\bs
\label{hhggff}
\ba
H &\equiv& c^d_1\, Y_{10}, \label{duiyp}
\label{hh} \\
F &\equiv& c^d_2\, Y_{\overline{126}}, \label{suiyp}
\label{ff} \\
G &\equiv& \sqrt{\left( c^d_5 \right)^2 + \left( c^d_6 \right)^2}\, Y_{120}.
\label{bnlhp}
\ea
\es
Then,
by using equations~(\ref{c}) and~\eqref{bnlhp} we derive
\bs
\label{bihop}
\ba
\left( c^d_5 \bar\alpha_5 + c^d_6 \bar\alpha_6 \right) Y_{120} 
&=&
\frac{c^d_5 \bar\alpha_5 + c^d_6 \bar\alpha_6}
{\sqrt{\left( c^d_5 \right)^2 + \left( c^d_6 \right)^2}}\ G
\no &=&
\left( \frac{\sqrt{3}}{2}\, \bar\alpha_5 + \frac{1}{2}\, \bar\alpha_6 \right) G,
\\
\left( c^\ell_5 \bar\alpha_5 + c^\ell_6 \bar\alpha_6 \right) Y_{120} 
&=&
\frac{c^\ell_5 \bar\alpha_5 + c^\ell_6 \bar\alpha_6}
{\sqrt{\left( c^\ell_5 \right)^2 + \left( c^\ell_6 \right)^2}}\
\sqrt{\frac{\left( c^\ell_5 \right)^2 +
\left( c^\ell_6 \right)^2}{\left( c^d_5 \right)^2 + \left( c^d_6 \right)^2}}\ G
\no &=&
\left( \frac{1}{2}\, \bar\alpha_5 - \frac{\sqrt{3}}{2}\, \bar\alpha_6 \right)
\sqrt{3}\, G,
\\
\left( c^u_5 \alpha_5 + c^u_6 \alpha_6 \right) Y_{120}
&=&
\frac{c^u_5 \alpha_5 + c^u_6 \alpha_6}
{\sqrt{\left( c^u_5 \right)^2 + \left( c^u_6 \right)^2}}\ 
\sqrt{\frac{\left( c^u_5 \right)^2 + \left( c^u_6 \right)^2}
{\left( c^d_5 \right)^2 + \left( c^d_6 \right)^2}}\ G
\no &=&
\left( - \frac{\sqrt{3}}{2}\, \alpha_5 + \frac{1}{2}\, \alpha_6 \right) G,
\\
\left( c^D_5 \alpha_5 + c^D_6 \alpha_6 \right) Y_{120}
&=&
\frac{c^D_5 \alpha_5 + c^D_6 \alpha_6}
{\sqrt{\left( c^D_5 \right)^2 + \left( c^D_6 \right)^2}}\
\sqrt{\frac{\left( c^D_5 \right)^2 + \left( c^D_6 \right)^2}
{\left( c^d_5 \right)^2 + \left( c^d_6 \right)^2}}\ G
\no &=&
\left( - \frac{1}{2}\, \alpha_5 - \frac{\sqrt{3}}{2}\, \alpha_6 \right)
\sqrt{3}\, G. 
\ea
\es
Equations~\eqref{hh},
\eqref{ff},
and~\eqref{bihop} may now be plugged into the mass formulas~(\ref{M}). 
The result is
\bs
\label{ftrid}
\ba
M_d &=& v \cos\beta \left[ \bar\alpha_1 H + \bar\alpha_2 F + 
\left( \frac{\sqrt{3}}{2}\, \bar\alpha_5 + \frac{1}{2}\, \bar\alpha_6 \right) G
\right],
\\
M_\ell &=& v \cos\beta \left[ \bar\alpha_1 H -3 \bar\alpha_2 F + 
\left( \frac{1}{2}\, \bar\alpha_5 - \frac{\sqrt{3}}{2}\, \bar\alpha_6 \right)
\sqrt{3}\, G
\right],
\\
M_u &=& v \sin\beta \left[ \alpha_1 H - \alpha_2 F +
\left( - \frac{\sqrt{3}}{2}\, \alpha_5 + \frac{1}{2}\, \alpha_6 \right) G
\right],
\\
M_D &=& v \sin\beta \left[ \alpha_1 H + 3 \alpha_2 F +
\left( - \frac{1}{2}\, \alpha_5 - \frac{\sqrt{3}}{2}\, \alpha_6 \right)
\sqrt{3}\, G \right].
\ea
\es
By comparing equations~\eqref{ftrid}
and~\eqref{mmm} we obtain the identifications
\bs
\ba
k_d &=& v \cos{\beta}\, \bar \alpha_1,
\\
k_u &=& v \sin{\beta}\, \alpha_1,
\\
v_d &=& v \cos{\beta}\, \bar \alpha_2,
\\
v_u &=& - v \sin{\beta}\, \alpha_2,
\\ \label{kappa_d}
\kappa_d &=& v \cos{\beta}
\left( \frac{\sqrt{3}}{2}\, \bar\alpha_5 + \frac{1}{2}\, \bar\alpha_6 \right),
\\ \label{kappa_ell}
\kappa_\ell &=& v \cos{\beta} 
\left( \frac{1}{2}\, \bar\alpha_5 - \frac{\sqrt{3}}{2}\, \bar\alpha_6 \right)
\sqrt{3},
\\ \label{kappa_u}
\kappa_u &=& v \sin{\beta}
\left( - \frac{\sqrt{3}}{2}\, \alpha_5 + \frac{1}{2}\, \alpha_6 \right),
\\ \label{kappa_D}
\kappa_D &=& v \sin{\beta}
\left( - \frac{1}{2}\, \alpha_5 - \frac{\sqrt{3}}{2}\, \alpha_6 \right)
\sqrt{3}.
\ea
\es
Computing $\bar\alpha_5$ and $\bar\alpha_6$ from equations~(\ref{kappa_d})
and~(\ref{kappa_ell}) gives 
\be
\left| \bar\alpha_5 \right|^2  + \left| \bar\alpha_6 \right|^2 = 
\frac{1}{v^2 \cos^2{\beta}} 
\left( \left| \kappa_d \right|^2
+ \frac{1}{3} \left| \kappa_\ell \right|^2 \right), 
\ee
while computing $\alpha_5$ and $\alpha_6$ from equations~(\ref{kappa_u})
and~(\ref{kappa_D}) leads to
\be
\left| \alpha_5 \right|^2 + \left| \alpha_6 \right|^2 =
\frac{1}{v^2 \sin^2\beta} 
\left( \left| \kappa_u \right|^2
+ \frac{1}{3} \left| \kappa_D \right|^2 \right).
\ee
Finally,
the consistency conditions~\eqref{alpha}
may be translated into the following conditions for the VEVs:
\bs
\label{vhpty}
\ba
\left| k_d \right|^2 + \left| v_d \right|^2
+ \left| \kappa_d \right|^2 + \frac{1}{3} \left| \kappa_\ell \right|^2
&\leq& v^2 \cos^2{\beta},
\\
\left| k_u \right|^2 + \left| v_u \right|^2
+ \left| \kappa_u \right|^2 + \frac{1}{3} \left| \kappa_D \right|^2
&\leq& v^2 \sin^2{\beta}.
\ea
\es

\end{appendix}

\vspace*{2mm}

\paragraph{Acknowledgements:}
We thank Ivo de Medeiros Varzielas for help
in the initial stages of the fitting procedure.
The work of W.G.~was partly supported
by a grant from the project ``NUPHYS 2013'' 20NIMLRD0287/CC 1140
of the \textit{Associa\c c\~ao do Instituto Superior T\'ecnico
para a Investiga\c c\~ao e Desenvolvimento} (IST-ID).
D.J.~thanks the Lithuanian Academy of Sciences
for support through the project~DaFi2015.
The work of L.L.~was supported through the projects PEst-OE-FIS-UI0777-2013,
PTDC/FIS-NUC/0548-2012,
and CERN-FP-123580-2011
of \textit{Funda\c c\~ao para a Ci\^encia e a Tecnologia} (FCT);
those projects are partially funded through POCTI (FEDER),
COMPETE,
QREN,
and the European Union.

\end{document}